\documentclass[aps,prb,onecolumn,notitlepage,groupedaddress,showpacs]{revtex4-1}

\usepackage{graphicx}

\begin{document}

% Use the \preprint command to place your local institutional report
% number in the upper righthand corner of the title page in preprint mode.
% Multiple \preprint commands are allowed.
% Use the 'preprintnumbers' class option to override journal defaults
% to display numbers if necessary
%\preprint{}

%Title of paper
\title{Fermi sea term in the relativistic linear 
       muffin-tin-orbital transport theory for random alloys}

% repeat the \author .. \affiliation  etc. as needed
% \email, \thanks, \homepage, \altaffiliation all apply to the current
% author. Explanatory text should go in the []'s, actual e-mail
% address or url should go in the {}'s for \email and \homepage.
% Please use the appropriate macro foreach each type of information

% \affiliation command applies to all authors since the last
% \affiliation command. The \affiliation command should follow the
% other information
% \affiliation can be followed by \email, \homepage, \thanks as well.
\author{I. Turek}
\email[]{turek@ipm.cz}
%\homepage[]{Your web page}
%\thanks{}
%\altaffiliation{}
\affiliation{Institute of Physics of Materials,
Academy of Sciences of the Czech Republic,
\v{Z}i\v{z}kova 22, CZ-616 62 Brno, Czech Republic}

\author{J. Kudrnovsk\'y}
\email[]{kudrnov@fzu.cz}
\affiliation{Institute of Physics, 
Academy of Sciences of the Czech Republic,
Na Slovance 2, CZ-182 21 Praha 8, Czech Republic}

\author{V. Drchal}
\email[]{drchal@fzu.cz}
\affiliation{Institute of Physics, 
Academy of Sciences of the Czech Republic,
Na Slovance 2, CZ-182 21 Praha 8, Czech Republic}

%Collaboration name if desired (requires use of superscriptaddress
%option in \documentclass). \noaffiliation is required (may also be
%used with the \author command).
%\collaboration can be followed by \email, \homepage, \thanks as well.
%\collaboration{}
%\noaffiliation

\date{\today}

\begin{abstract}
We present a formulation of the so-called Fermi sea contribution
to the conductivity tensor of spin-polarized random alloys within
the fully relativistic tight-binding linear muffin-tin-orbital
(TB-LMTO) method and the coherent potential approximation (CPA).
We show that the configuration averaging of this contribution 
leads to the CPA-vertex corrections that are solely due to the
energy dependence of the average single-particle propagators.
Moreover, we prove that this contribution is indispensable for the
invariance of the anomalous Hall conductivities with respect to the
particular LMTO representation used in numerical implementation.
\emph{Ab initio} calculations for cubic ferromagnetic $3d$ transition
metals (Fe, Co, Ni) and their random binary alloys (Ni-Fe, Fe-Si)
indicate that the Fermi sea term is small against the dominating Fermi
surface term.
However, for more complicated structures and systems, such as
hexagonal cobalt and selected ordered and disordered Co-based Heusler
alloys, the Fermi sea term plays a significant role in the
quantitative theory of the anomalous Hall effect.
\end{abstract}

% insert suggested PACS numbers in braces on next line
% insert suggested keywords - APS authors don't need to do this
%\keywords{}

%\maketitle must follow title, authors, abstract, \pacs, and \keywords
\maketitle

% body of paper here - Use proper section commands
% References should be done using the \cite, \ref, and \label commands
% Put \label in argument of \section for cross-referencing
%\subsection{}
%\subsubsection{}

\section{Introduction\label{s_intr}}

The simultaneous presence of a spontaneous spin polarization and
spin-orbit interaction in a solid gives rise to a number of
physically interesting and technologically important phenomena, such
as, e.g., magnetocrystalline anisotropy or magnetic dichroism in the
X-ray absorption spectra. \cite{r_1998_ps}
The anomalous Hall effect \cite{r_1881_eh, r_2008_nas, r_2010_nso}
(AHE) represents the most famous example of a spin-orbit driven
transverse transport phenomenon in itinerant magnets; 
these phenomena include also the anomalous Nernst effect
\cite{r_2006_xyf, r_2008_osn} as a thermal analog of the AHE.
The current understanding of the AHE rests on the identification of the
basic underlying mechanisms, \cite{r_2001_cb, r_2008_nas, r_2010_nso} 
namely, the Berry curvature of occupied Bloch states for perfect
crystals \cite{r_1954_kl, r_2002_jnm} and the skew scattering
\cite{r_1955_js, r_1958_js} and side-jump \cite{r_1970_lb} mechanisms
for systems with impurities.
The present state of the theory of the AHE can be documented by ample 
model studies explaining the observed relation between the
anomalous Hall conductivity (AHC) and the longitudinal conductivity
in different regimes, ranging from low to high conductivities, 
see Ref.~\onlinecite{r_2010_nso, r_2008_osn} and references therein.

Recently, first-principles studies of specific materials have 
appeared, devoted to pure ferromagnetic 
metals \cite{r_2004_ykm, r_2007_wvy, r_2009_rms} and various ordered
compounds, \cite{r_2003_ivs, r_2012_kf, r_2013_tg} including
spin-gapless semi\-conductors \cite{r_2013_off} and non-collinear
antiferromagnets. \cite{r_2014_cnm}
The latest trend in this field is featured by a development of
techniques applicable to a wide class of systems, covering both
clean crystals and diluted as well as concentrated random alloys,
with the inclusion of all contributions to the AHE on equal footing. 
Existing studies are based on the single-particle Green's function
(GF) and the coherent potential approximation (CPA) in the framework
of the fully relativistic Korringa-Kohn-Rostoker (KKR)
\cite{r_2010_lke} or the tight-binding linear muffin-tin-orbital 
(TB-LMTO) \cite{r_2012_tkd} methods.
Both approaches employ the conductivity tensor formulated in the
Kubo linear response theory, \cite{r_1957_rk} where the
corresponding configuration averaging leads to the so-called
CPA-vertex corrections. \cite{r_1969_bv, r_1985_whb, r_2006_ctk}

Most of the published results for random alloys \cite{r_2010_lke,
r_2013_kcm, r_2011_kdk, r_2012_tkd, r_2013_kdt} rest on the
Kubo-St\v{r}eda formula, \cite{r_1982_ps, r_2001_cb} which provides
an expression for the full conductivity tensor at zero temperature
solely in terms of quantities at the Fermi energy.
However, the calculations could be performed only with the neglect
of a term containing -- in addition to the usual velocity operators
-- also the coordinate operator.
This last operator is not compatible with periodic boundary conditions
used in standard bulk techniques; the neglect of the problematic term
has been justified by a high degree of symmetry of the crystal lattice
(e.g., cubic). \cite{r_2010_lke, r_2012_tkd} 
The neglected term is equivalent to the so-called Fermi sea
contribution which follows from the original Bastin formula for
the conductivity tensor. \cite{r_1971_blb}
The Fermi sea term does not contain any coordinate operator, but it
requires energy integration over the occupied part of the valence
spectrum.
Its direct evaluation for model systems \cite{r_2007_kty} and for
realistic band structures of transition metals \cite{r_2010_nhk}
indicates that, at least in the high-conductivity regime, the Fermi
sea term represents a small correction to the dominating Fermi
surface term defined only in terms of quantities at the Fermi energy.
However, similar studies for the qualitatively different systems
mentioned above cannot be found in the literature.

The present study is devoted to a formulation of the conductivity
tensor from the Bastin formula in the relativistic TB-LMTO method;
the focus is on the Fermi sea term and random alloys.
Our approach employs several differences between the TB-LMTO and
KKR methods.
First, the transport in the TB-LMTO method is described as intersite
hopping between neighboring atomic (Wigner-Seitz) cells, which leads
to non-random (configuration independent) effective velocity
operators, \cite{r_2002_tkd} while the KKR method rests on velocities
represented by random site-diagonal matrices. \cite{r_1985_whb}
This feature is advantageous especially in the formulation of the
vertex corrections in the TB-LMTO-CPA approach.
Second, the LMTO structure constants are energy independent in
contrast to the energy dependent KKR structure constants.
Third, the scattering properties of individual atoms are described
by simple analytic functions of energy in the LMTO method, whereas
the full energy dependence of the single-site T-matrices in the KKR
method is obtained by numerical integration of the radial 
Schr\"odinger or Dirac equation. \cite{r_2005_zhs, r_2011_ekm} 
The last two properties are used in the calculation of the energy
derivative of the GF which enters the Fermi sea term.

The paper is organized as follows.
The theoretical part is summarized in Section \ref{s_meth},
with Section \ref{ss_ctbf} devoted to the general form
of the conductivity tensor and Section \ref{ss_cafst}
explaining the configuration averaging of the Fermi sea term.
The technical aspects of the energy derivative of the coherent
potential function, appearing in the final formula, are
presented in Appendix \ref{app_ed}.
The LMTO transformation properties of individual parts of the
derived conductivity tensor are listed in Section \ref{ss_tpct}
while the corresponding details can be found in Appendix
\ref{app_tict}.
Information on the numerical implementation is given in
Section \ref{ss_ind}.
Illustrating examples of the calculated AHCs are summarized
in Section \ref{s_redi}; the case of pure metals is contained
in Section \ref{ss_pfcn} and that of random alloys and compounds
is discussed in Section \ref{ss_rac}.
The main conclusions are presented in Section \ref{s_conc}.

\section{Method\label{s_meth}}

\subsection{Conductivity tensor from 
            the Bastin formula\label{ss_ctbf}} 

Our starting point is the so-called Bastin formula for the
conductivity tensor $\sigma_{\mu\nu}$ of a non-interacting electron
system 
\begin{eqnarray}
\sigma_{\mu\nu} & = & -2\sigma_0 \int \mathrm{d}E f(E) \mathrm{Tr}
\left\langle  V_\mu G'_+(E) V_\nu [ G_+(E) - G_-(E) ] \right.
\nonumber\\
 & & \left. \qquad \qquad \qquad \qquad
 - V_\mu [ G_+(E) - G_-(E) ] V_\nu  G'_-(E) \right\rangle ,
\label{eq_bastin}
\end{eqnarray}
which is well documented in the literature. \cite{r_1971_blb,
r_1977_ss, r_1982_ps, r_2001_cb} 
The subscripts $\mu$ and $\nu$ in (\ref{eq_bastin}) are indices
of Cartesian coordinates ($\mu, \nu \in \{ x , y, z \}$),
the integral is taken over
the whole real energy axis, the function $f(E)$ denotes the
Fermi-Dirac distribution function, the symbol $V_\mu$ denotes the
velocity operator and the $G_\pm(E) = \lim_{\epsilon\to 0^+}
 G(E\pm\mathrm{i}\epsilon)$ denote the side limits of the GF
(resolvent) of the one-particle Hamiltonian $H$,
\begin{equation}
G(z) = ( z - H )^{-1} ,
\label{eq_defgz}
\end{equation}
where $z$ is a complex energy variable.
The prime at the $G_\pm(E)$ denotes the energy derivative and the
prefactor $\sigma_0$ is given by $\sigma_0 = e^2 \hbar/(4\pi V_0 N)$,
where $e$ denotes the electron charge,
$V_0$ is the volume of the primitive cell, and $N$ is the number of
cells in a big finite crystal with periodic boundary conditions. 
The trace in (\ref{eq_bastin}) is taken over all orbitals of the
Hilbert space of the crystal and the symbol $\langle \dots \rangle$
denotes the average over all configurations of the random alloy.
The velocity operators (with $\hbar = 1$ assumed) are defined by a
quantum-mechanical commutation relation as
\begin{equation}
V_\mu = - {\rm i} [ X_\mu , H ] ,
\label{eq_veloc}
\end{equation}
where $X_\mu$ denotes the coordinate operator.

The Hamiltonian in the relativistic TB-LMTO method for spin-polarized
systems can be written in the basis of orthonormal LMTO orbitals
as \cite{r_1991_slg, r_1996_sdk, r_1997_tdk, r_2012_tkd}
\begin{equation}
H = C + (\sqrt{\Delta})^+ \left[ 1 + S (\alpha - \gamma) \right]^{-1}
 S \sqrt{\Delta} ,
\label{eq_defh}
\end{equation}
where the symbols $C$, $\sqrt{\Delta}$ and $\gamma$ denote  
site-diagonal matrices of the LMTO potential parameters while $\alpha$
denotes a site-diagonal matrix of the LMTO screening constants, which
define both the corresponding LMTO representation and the matrix $S$
of short-ranged (screened) structure constants. \cite{r_1984_aj, 
r_1986_apj} 
To simplify the formulas below, the index $\alpha$ of the
particular LMTO representation is omitted in the main text of the
paper including Appendix \ref{app_ed}, but it is restored in 
Appendix \ref{app_tict}.

The TB-LMTO method for disordered alloys \cite{r_1990_kd, r_1997_tdk,
r_1996_sdk} employs the so-called auxiliary GF
\begin{equation}
g(z) = [ P(z) - S ]^{-1} ,
\label{eq_defga}
\end{equation}
where $P(z)$ denotes a site-diagonal matrix of potential functions.
Note that the potential functions $P(z)$ as well as the potential
parameters $C$, $\sqrt{\Delta}$ and $\gamma$ are random quantities,
depending on the occupation of the lattice sites ${\bf R}$ by atomic
species of the alloy, whereas the structure-constant matrix $S$ and
the matrix $\alpha$ of screening constants are non-random.
The configuration average of the $g(z)$ in the single-site CPA is
given by
\begin{equation}
\langle g(z) \rangle = {\bar g}(z) = [ \mathcal{P}(z) - S ]^{-1} ,
\label{eq_avaux}
\end{equation}
where $\mathcal{P}(z)$ denotes a non-random site-diagonal matrix
of the so-called coherent potential functions.

The TB-LMTO transport studies \cite{r_2002_tkd, r_2012_tkd} rest on
a systematic disregarding of electron motion inside the atomic
spheres, which leads to non-random effective velocities defined by
\begin{equation}
v_\mu = - {\rm i} [ X_\mu , S ] .
\label{eq_veloce}
\end{equation}
The coordinate operators in (\ref{eq_veloce}) are represented
by site- and orbital-diagonal matrices
\begin{equation}
( X_\mu )^{LL'}_{{\bf R}{\bf R}'} = \delta_{{\bf R}{\bf R}'}
\delta^{LL'} X^\mu_{\bf R} ,
\label{eq_coord}
\end{equation}
where $X^\mu_{\bf R}$ is the $\mu$th component of the position
vector ${\bf R}$ and $L$ denotes the orbital index. 
Note that the index $L$ in this paper labels all orbitals 
belonging to a single site; its detailed structure in the
spin-polarized relativistic formalism has been given elsewhere.
\cite{r_1996_sdk, r_1997_tdk, r_2012_tkd}

It was shown in our previous study \cite{r_2012_tkd} that the
original velocity (\ref{eq_veloc}) is related to its effective
counterpart (\ref{eq_veloce}) by a simple rescaling 
\begin{equation}
V_\mu = (\sqrt{\Delta})^+ F^{-1} \, v_\mu \, 
( F^+ )^{-1} \sqrt{\Delta} ,
\label{eq_velop}
\end{equation}
where the factor $F$ is explicitly given by
\begin{equation}
F = 1 + S (\alpha - \gamma) .
\label{eq_falpha}
\end{equation}
Similarly, the relation between the true resolvent (\ref{eq_defgz})
and the auxiliary GF (\ref{eq_defga}) is given by a matrix shift and 
rescaling \cite{r_2012_tkd}
\begin{equation}
G(z) = (\sqrt{\Delta})^{-1} F^+ \left[ (\alpha - \gamma)
 + g(z) F \right] [(\sqrt{\Delta})^+]^{-1} .
\label{eq_agvga}
\end{equation}
Since the quantities $\alpha$, $\gamma$, $\sqrt{\Delta}$ and $F$
are energy independent, the last relation simplifies for energy
derivatives of the GFs to 
\begin{equation}
G'(z) = (\sqrt{\Delta})^{-1} F^+ \, g'(z) \, F 
[(\sqrt{\Delta})^+]^{-1} ,
\label{eq_agvgap}
\end{equation}
and for differences of the GFs to
\begin{equation}
G(z_1) - G(z_2) = (\sqrt{\Delta})^{-1} F^+ 
\left[ g(z_1) - g(z_2) \right] F [(\sqrt{\Delta})^+]^{-1} ,
\label{eq_agvgad}
\end{equation}
where $z$, $z_1$ and $z_2$ are arbitrary complex energies outside
the spectrum of the Hamiltonian $H$.
The use of Eqs.~(\ref{eq_velop}, \ref{eq_agvgap}, \ref{eq_agvgad})
in the Bastin formula (\ref{eq_bastin}) leads immediately to
the following expression for the conductivity tensor:
\begin{eqnarray}
\sigma_{\mu\nu} & = & -2\sigma_0 \int \mathrm{d}E f(E) \mathrm{Tr}
\left\langle  v_\mu g'_+(E) v_\nu [ g_+(E) - g_-(E) ] \right.
\nonumber\\
 & & \left. \qquad \qquad \qquad \qquad
 - v_\mu [ g_+(E) - g_-(E) ] v_\nu  g'_-(E) \right\rangle .
\label{eq_bastin0}
\end{eqnarray}
This form is identical with the original one, but the derived
Eq.~(\ref{eq_bastin0}) has clear advantages in the configuration
averaging for two reasons. 
First, the full resolvents $G_\pm(E)$ are replaced by the auxiliary
GFs for which the CPA-average $\bar{g}(z)$ can be directly
evaluated according to Eq.~(\ref{eq_avaux}).
Second, the random velocities $V_\mu$ are replaced by the
non-random effective velocities $v_\mu$ so that the configuration
average of the whole conductivity tensor can be performed by 
following the standard formulation of the CPA-vertex 
corrections. \cite{r_1969_bv, r_2006_ctk}

Further processing of the expression (\ref{eq_bastin0}) can be done
along the well-known direction: \cite{r_2001_cb} one half of this
expression is kept while the integration of the second half is
performed by parts.
The result is then rewritten as a sum of two terms,
\begin{equation}
\sigma_{\mu\nu} = \sigma^{(1)}_{\mu\nu} + \sigma^{(2)}_{\mu\nu} ,
\label{eq_sigma}
\end{equation}
where the first term -- the Fermi surface term -- contains the
integral with $f'(E)$, namely,
\begin{eqnarray}
\sigma^{(1)}_{\mu\nu} & = & \sigma_0 \int \mathrm{d}E f'(E) 
\mathrm{Tr}
\left\langle  v_\mu g_+(E) v_\nu [ g_+(E) - g_-(E) ] \right.
\nonumber\\
 & & \left. \qquad \qquad \quad \quad \ \,
 - v_\mu [ g_+(E) - g_-(E) ] v_\nu  g_-(E) \right\rangle ,
\label{eq_surfi}
\end{eqnarray}
while the second term -- the Fermi sea term -- contains the
rest integral with $f(E)$, i.e.,
\begin{eqnarray}
\sigma^{(2)}_{\mu\nu} & = & \sigma_0 \int \mathrm{d}E f(E) 
\mathrm{Tr} \left\langle 
  v_\mu g_+(E) v_\nu g'_+(E) 
- v_\mu g'_+(E) v_\nu g_+(E) \right.
\nonumber\\
 & & \left. \qquad \qquad \quad \quad \ \,
 - v_\mu g_-(E) v_\nu g'_-(E) 
 + v_\mu g'_-(E) v_\nu g_-(E) \right\rangle .
\label{eq_seai}
\end{eqnarray}
For systems with zero temperature, the Fermi surface term 
(\ref{eq_surfi}) can be written as
\begin{eqnarray}
\sigma^{(1)}_{\mu\nu} & = & \sigma_0 \mathrm{Tr} 
\left\langle  v_\mu [ g_+(E_\mathrm{F}) - g_-(E_\mathrm{F}) ] 
v_\nu  g_-(E_\mathrm{F}) \right.
\nonumber\\
 & & \left. \qquad 
 - v_\mu g_+(E_\mathrm{F}) v_\nu [ g_+(E_\mathrm{F}) 
 - g_-(E_\mathrm{F}) ] \right\rangle ,
\label{eq_surf}
\end{eqnarray}
where $E_\mathrm{F}$ denotes the Fermi energy. 
The evaluation of this term within the relativistic LMTO approach
has been described in Ref.~\onlinecite{r_2012_tkd} while the
corresponding vertex corrections were formulated in detail in
the Appendix of Ref.~\onlinecite{r_2006_ctk}.

The zero-temperature case of the Fermi sea term (\ref{eq_seai}) can
be recast into a contour integral in the complex energy plane:
\begin{equation}
\sigma^{(2)}_{\mu\nu} = \sigma_0 \int_C \mathrm{d}z  
\mathrm{Tr} \left\langle v_\mu g'(z) v_\nu g(z) 
- v_\mu g(z) v_\nu g'(z) \right\rangle ,
\label{eq_seaj}
\end{equation}
where the integration path $C$ starts and ends at $E_\mathrm{F}$,
it is oriented counterclockwise and it encompasses the whole occupied
part of the alloy valence spectrum.
Note that the Fermi sea term is antisymmetric, $\sigma^{(2)}_{\mu\nu}
 = - \sigma^{(2)}_{\nu\mu}$, so that it contributes only to the
AHCs while the longitudinal conductivities are given only by the Fermi
surface term.

The derived expression (\ref{eq_seaj}) for the Fermi sea term,
obtained from the Kubo-Bastin formulation of the conductivity tensor,
can be transformed into the corresponding term of the Kubo-St\v{r}eda
formula, namely, $\sigma^{(2)}_{\mu\nu} = \sigma_0 \mathrm{Tr} 
\langle \mathrm{i} ( X_\mu v_\nu - X_\nu v_\mu ) 
[ g_+(E_\mathrm{F}) - g_-(E_\mathrm{F}) ] \rangle$, 
see the second term in Eq.~(17) of Ref.~\onlinecite{r_2012_tkd}. 
This transformation is formally exact, \cite{r_1982_ps, r_2001_cb}
but the transformed result contains the coordinate operator that is
unbounded and incompatible with the periodic boundary conditions used
in the numerical implementation. 
For simple systems with inversion symmetry (cubic, hexagonal
close-packed), the lattice summations can be rearranged in such a way
that the resulting $\sigma^{(2)}_{\mu\nu}$ vanishes identically, in
contrast to the results of Eq.~(\ref{eq_seaj}), see 
Section~\ref{ss_pfcn}.
Since the Kubo-Bastin approach does not involve the problematic
coordinate operator and since the direct evaluation of 
$\sigma^{(2)}_{\mu\nu}$ yields non-zero values even for the simplest
systems, \cite{r_2007_kty, r_2010_nhk} the derived formula
(\ref{eq_seaj}) represents a correct version of the Fermi sea term
within the TB-LMTO formalism.

\subsection{Configuration averaging of the Fermi sea
            term\label{ss_cafst}} 

For the CPA-average of the Fermi sea term (\ref{eq_seaj}), we use the
relation
\begin{equation}
\mathrm{Tr} \langle v_\mu g'(z) v_\nu g(z) \rangle =
\lim_{z_1 \to z} \frac{\partial}{\partial z_1} 
\mathrm{Tr} \langle v_\mu g(z_1) v_\nu g(z) \rangle ,
\label{eq_vgpvgi}
\end{equation}
where the average on the r.h.s.\ can be split into the coherent
part and the incoherent part (vertex corrections -- VC):
\begin{equation}
\mathrm{Tr} \langle v_\mu g(z_1) v_\nu g(z) \rangle =
\mathrm{Tr} \{ v_\mu \bar{g}(z_1) v_\nu \bar{g}(z) \} +
\mathrm{Tr} \langle v_\mu g(z_1) v_\nu g(z) \rangle_\mathrm{VC} .
\label{eq_vg1vg}
\end{equation}
The second term can be written according to the general
expression \cite{r_2006_ctk} as
\begin{eqnarray}
\mathrm{Tr} \langle v_\mu g(z_1) v_\nu g(z) \rangle_\mathrm{VC} 
& = & \sum_{{\bf R}_1 \Lambda_1} \sum_{{\bf R}_2 \Lambda_2} 
\left[ \bar{g}(z) v_\mu \bar{g}(z_1) 
\right]^{L'_1 L_1}_{{\bf R}_1 {\bf R}_1}
\left[ \Delta^{-1}( z_1 , z ) 
\right]^{\Lambda_1 \Lambda_2}_{{\bf R}_1 {\bf R}_2}
\nonumber\\
 & & \qquad \quad \times
\left[ \bar{g}(z_1) v_\nu \bar{g}(z) 
\right]^{L_2 L'_2}_{{\bf R}_2 {\bf R}_2} ,
\label{eq_vg1vgvc}
\end{eqnarray}
where the symbols $\Lambda_1$ and $\Lambda_2$ abbreviate
the composed orbital indices $\Lambda_1 = ( L_1 , L'_1 )$ and
$\Lambda_2 = ( L_2 , L'_2 )$ and where the matrix 
$\Delta^{\Lambda_1 \Lambda_2}_{{\bf R}_1 {\bf R}_2}( z_1 , z_2 )$
was defined in the Appendix of Ref.~\onlinecite{r_2006_ctk}.
The evaluation of the vertex contribution to Eq.~(\ref{eq_vgpvgi})
is greatly simplified due to the exact vanishing of the on-site
blocks of the matrix product $\bar{g}(z) v_\mu \bar{g}(z)$:  
\begin{equation}
[ \bar{g}(z) v_\mu \bar{g}(z) ]^{LL'}_{{\bf R}{\bf R}} = 0 ,
\label{eq_gvgon}
\end{equation}
which is valid for the same energy arguments of both GFs.
This rule is a consequence of the simple form of the coordinate
operators $X_\mu$ (\ref{eq_coord}) and of the single-site nature
of the coherent potential functions $\mathcal{P}(z)$, i.e.,
$\mathcal{P}^{LL'}_{{\bf R}{\bf R}'}(z) = 
\delta_{{\bf R}{\bf R}'} \mathcal{P}^{LL'}_{\bf R}(z)$, from which
we get $[ \mathcal{P}(z) , X_\mu ] = 0$ and, by employing
Eq.~(\ref{eq_avaux}), also $\bar{g}(z) [ X_\mu , S ] \bar{g}(z)
= \bar{g}(z) [ \mathcal{P}(z) - S , X_\mu ] \bar{g}(z) = 
[ X_\mu , \bar{g}(z) ]$. 
The validity of Eq.~(\ref{eq_gvgon}) follows now from the vanishing
of the on-site blocks of the last commutator.
After taking the partial derivative with respect to $z_1$ of
Eq.~(\ref{eq_vg1vgvc}), making the limit $z_1 \to z$, and using
the rule (\ref{eq_gvgon}), we get
\begin{equation}
\lim_{z_1 \to z} \frac{\partial}{\partial z_1} 
\mathrm{Tr} \langle v_\mu g(z_1) v_\nu g(z) \rangle_\mathrm{VC}
= 0 .
\label{eq_vcvan}
\end{equation}
By employing this identity in Eqs.~(\ref{eq_vgpvgi}, \ref{eq_vg1vg}),
we obtain a simple result
\begin{equation}
\mathrm{Tr} \langle v_\mu g'(z) v_\nu g(z) \rangle =
\mathrm{Tr} \{ v_\mu \bar{g}'(z) v_\nu \bar{g}(z) \} ,
\label{eq_vgpvgf}
\end{equation}
which can be used to rewrite the configurationally averaged Fermi
sea term (\ref{eq_seaj}) as
\begin{equation}
\sigma^{(2)}_{\mu\nu} = \sigma_0 \int_C \mathrm{d}z  
\mathrm{Tr} \left\{ v_\mu \bar{g}'(z) v_\nu \bar{g}(z) 
- v_\mu \bar{g}(z) v_\nu \bar{g}'(z) \right\} .
\label{eq_seaq}
\end{equation}
This formula represents the main result of this Section.

The obtained result (\ref{eq_seaq}) could be interpreted as if the
averaged Fermi sea contribution to the conductivity tensor contained
only the coherent part. 
However, in the practical evaluation of the energy derivative 
$\bar{g}'(z)$, one has to use the relation 
\begin{equation}
\bar{g}'(z) = - \bar{g}(z) \mathcal{P}'(z) \bar{g}(z) ,
\label{eq_bgp}
\end{equation}
which follows from the energy independent structure constants $S$
in (\ref{eq_avaux}) and which leads to the final expression for the
Fermi sea term: 
\begin{equation}
\sigma^{(2)}_{\mu\nu} = \sigma_0 \int_C \mathrm{d}z  
\mathrm{Tr} \left\{ [ v_\mu \bar{g}(z) v_\nu 
- v_\nu \bar{g}(z) v_\mu ] \bar{g}(z)  
\mathcal{P}'(z) \bar{g}(z) \right\} .
\label{eq_seaf}
\end{equation}
As it is shown in Appendix \ref{app_ed}, the formulation of the energy
derivative $\mathcal{P}'(z)$ leads to a set of linear equations that
is very similar to that encountered in the formulation of the
CPA-vertex corrections. \cite{r_2006_ctk}
In other words, in the configuration average of the Fermi sea term,
the vertex corrections corresponding to $\langle g(z) v_\mu g(z) 
\rangle$ vanish identically, but those appearing in $\bar{g}'(z) =
 - \langle g(z) P'(z) g(z) \rangle$ do contribute.
These last vertex corrections are related directly to the energy
dependence of the averaged single-particle GF and to the Ward identity
for the conservation of particle number; \cite{r_1969_bv} 
their proper inclusion is thus inevitable for an internally
consistent approximative theory of the conductivity tensor.

Let us discuss briefly properties of the Fermi sea term in the
dilute limit of a random binary alloy A$_{1-c}$B$_c$, where
$c \to 0^+$.
The AHC exhibits a divergent behavior in this limit, that is due to
the incoherent part (vertex corrections) of the Fermi surface
term. \cite{r_2010_lke}
The coherent potential function behaves for small concentrations
$c$ as \cite{r_1968_vke}
\begin{equation}
\mathcal{P}_{\bf R}(z) = 
P^\mathrm{A}_{\bf R}(z) + c t^\mathrm{B}_{\bf R}(z) ,
\label{eq_cpfdl}
\end{equation}
where it is assumed that the species-resolved potential functions
$P^\mathrm{A}_{\bf R}(z)$ and $P^\mathrm{B}_{\bf R}(z)$ are
concentration independent and where the $t^\mathrm{B}_{\bf R}(z)$
denotes the single-site T-matrix of a single B impurity in the
host A crystal, see Eq.~(\ref{eq_tmat}) with 
$\mathcal{P}_{\bf R}(z) = P^\mathrm{A}_{\bf R}(z)$ 
and $P_{\bf R}(z) = P^\mathrm{B}_{\bf R}(z)$.
This regular concentration dependence of the $\mathcal{P}_{\bf R}(z)$
and the final form of $\sigma^{(2)}_{\mu\nu}$ (\ref{eq_seaf}) mean
that the Fermi sea term behaves in general regularly in the
dilute limit.
The only exceptions to this rule might be due to a possible
singularity of the impurity T-matrix $t^\mathrm{B}_{\bf R}(z)$ at the
Fermi energy.
The limiting case of a clean A crystal ($c=0$) is obtained by setting
$\mathcal{P}_{\bf R}(z) = P^\mathrm{A}_{\bf R}(z)$ in 
Eq.~(\ref{eq_seaf}) as well as in all GFs; the total AHC is then
finite and equivalent to that of the Berry-curvature approach.

\subsection{Transformation properties of the conductivity
            tensor\label{ss_tpct}} 

The TB-LMTO method for perfect crystals can be formulated in
a general LMTO representation specified by a set of site-diagonal
screening constants. \cite{r_1984_aj, r_1986_apj}
Most auxiliary quantities depend on the choice of the LMTO
representation and they have to be transformed according to
well-known relations when changing the LMTO representation, whereas
all physical quantities remain invariant.
These transformation properties have been successfully combined with
the CPA for one-particle quantities of random alloys, such as the
average auxiliary GF $\bar{g}(z)$, the coherent potential function
$\mathcal{P}(z)$, or the single-site T-matrices 
$t_{\bf R}(z)$. \cite{r_1997_tdk, r_2000_tkd}
However, the case of two-particle quantities, in particular of the
conductivity tensor, has not been treated so far.
Here we summarize the most important results concerning the total
tensor and its various contributions; the proof of our statements is
outlined in Appendix \ref{app_tict}.

Let us write the conductivity tensor (\ref{eq_sigma}) as
$\sigma_{\mu\nu} = \sigma^{(1)}_{\mu\nu,\mathrm{coh}} +
\sigma^{(1)}_{\mu\nu,\mathrm{VC}} + \sigma^{(2)}_{\mu\nu}$,
where the first and the second terms denote, respectively, the
coherent and the incoherent (vertex) parts of the Fermi surface term
(\ref{eq_surf}). 
The following quantities are then invariant:
(i) the total tensor $\sigma_{\mu\nu}$, 
(ii) the incoherent Fermi surface term
 $\sigma^{(1)}_{\mu\nu,\mathrm{VC}}$, and
(iii) the sum of the coherent Fermi surface term and of the
Fermi sea term, $\sigma^{(1)}_{\mu\nu,\mathrm{coh}} + 
\sigma^{(2)}_{\mu\nu}$. 
Since the Fermi sea term is antisymmetric, the last property means
that the symmetric part of the coherent Fermi surface term, 
$[\sigma^{(1)}_{\mu\nu,\mathrm{coh}} + 
  \sigma^{(1)}_{\nu\mu,\mathrm{coh}}]/2$, is invariant as well.
These properties prove the importance of the Fermi sea term for
the complete TB-LMTO-CPA theory of the AHE.

The above LMTO transformation properties together with the purely
coherent nature of the Fermi sea term (\ref{eq_seaq}) and with its
regular behavior in diluted alloys (end of Section \ref{ss_cafst})
are relevant for a classification of the intrinsic and extrinsic
contributions to the AHE. \cite{r_2008_nas, r_2010_nso, r_2010_lke}
Within the present TB-LMTO formalism, the intrinsic AHC has to be
identified with the antisymmetric part of the sum of the coherent
Fermi surface term and of the Fermi sea term 
[$\sigma^{(1)}_{\mu\nu,\mathrm{coh}} + \sigma^{(2)}_{\mu\nu}$],
whereas the extrinsic AHC is given by the antisymmetric part of the 
incoherent Fermi surface term [$\sigma^{(1)}_{\mu\nu,\mathrm{VC}}$].
This seems to be a natural generalization of the classification
introduced recently in the KKR method using the Kubo-St\v{r}eda
formula. \cite{r_2010_lke, r_2013_kcm} 

\subsection{Implementation and numerical details\label{ss_ind}} 

The numerical implementation of the developed scheme and the
calculations discussed in Section \ref{s_redi} were done
with similar parameters as described in our recent papers
concerning both the fully relativistic selfconsistent electronic
structures \cite{r_2012_tkc} and the Fermi surface term of the
conductivity tensor. \cite{r_2012_tkd, r_2013_kdt}
The particular LMTO representation used in the calculations is
defined by the screening constants leading to the most localized
real-space structure constants for the valence basis consisting of
$s$-, $p$-, and $d$-type orbitals. \cite{r_1984_aj, r_1986_apj}
For calculations of the Fermi surface term, a small imaginary
part of $\pm 10^{-5}$ Ry has been added to the Fermi energy while
in the evaluation of the Fermi sea term (\ref{eq_seaf}), the
integration has been performed along a circular contour of a
diameter 1.5 Ry. \cite{r_1982_wfl, r_1982_zdd}
The contour integral was approximated by a sum over 20 -- 40 complex
nodes in the upper semicircle; the nodes were located in an
asymmetric way which results in a denser mesh near the Fermi energy.
The number of ${\bf k}$ vectors sampling the Brillouin zone 
depends on the distance between the particular complex node and
the Fermi energy; for the node closest to the Fermi energy, the total
numbers of $\sim 10^8$ ${\bf k}$ vectors have been used.
Convergence tests with respect to the numbers of energy nodes and
of ${\bf k}$ vectors have been performed for each system, 
which guarantee the reliability of the results presented below.
In the present study, we have included three shells of nearest
neighbors in the screened structure-constant matrix of the
body-centered cubic (bcc) lattice, in contrast to the two neighboring
shells used in Refs.~\onlinecite{r_2012_tkd, r_2013_kdt}, which leads
to slightly modified values of the Fermi surface terms.
Moreover, in contrast to our previous studies, the sign convention
of the AHC in the present study has been adopted according to other
authors. \cite{r_2004_ykm, r_2012_kf, r_2010_lke}

\section{Results and discussion\label{s_redi}}

\subsection{Pure Fe, Co, and Ni\label{ss_pfcn}}

The calculated AHCs for pure Fe, Co and Ni, compared to the results
of other authors and with measured low-temperature values, 
\cite{r_1967_pnd, r_2012_hlw, r_2012_ytj} are shown in 
Table \ref{t_fcn}.
For the cubic metals, bcc Fe and face-centered cubic (fcc) Co and Ni,
the magnetization direction was taken along the [001] direction while
for hexagonal close-packed (hcp) Co, magnetization was considered
pointing along the hexagonal $c$ axis (easy axis) as well as lying
in the $ab$ plane perpendicular to it.
Note that the theoretical approaches based on the Berry curvature
\cite{r_2004_ykm, r_2007_wvy, r_2009_rms} include both the Fermi
surface and the Fermi sea terms, whereas the published KKR results
in Ref.~\onlinecite{r_2013_kcm} contain only the Fermi surface term.

\begin{table}
\caption{The calculated and experimental values of the AHC 
(in S/cm) for ferromagnetic $3d$ transition metals.
Two columns for hcp Co refer to the magnetization direction along
the $c$ axis ($c$) and in the $ab$ plane ($ab$).
The values of the Fermi sea term are displayed in parentheses.
\label{t_fcn}}
\begin{ruledtabular}
\begin{tabular}{lccccc}
       & bcc Fe & hcp Co ($c$) & hcp Co ($ab$) & fcc Co & fcc Ni \\
\hline
This work & 
 $796$ ($179$) & $471$ ($181$) & $169$ ($66$) & 
 $359$ ($-5$) & $-2432$ ($-17$) \\
Berry curvature & 
 $751$\footnote{Reference \onlinecite{r_2004_ykm}.} & 
 $481$\footnote{Reference \onlinecite{r_2009_rms}.} & 
 $116^{\text{b}}$ & $249^{\text{b}}$ & 
 $-2203$\footnote{Reference \onlinecite{r_2007_wvy}.} \\
KKR method & 
 $685$\footnote{Reference \onlinecite{r_2013_kcm}.} & 
 $325^{\text{d}}$ & & $213^{\text{d}}$ & $-2062^{\text{d}}$ \\
Experiment & 
 $1032$\footnote{Reference \onlinecite{r_1967_pnd}.} &
 $\sim 813^{\text{b}}$ & $\sim 150^{\text{b}}$ & 
 $727$\footnote{Reference \onlinecite{r_2012_hlw}.} & 
 $-1100$\footnote{Reference \onlinecite{r_2012_ytj}.} \\
\end{tabular}
\end{ruledtabular}
\end{table}

The total AHCs calculated in this work are in reasonable agreement
with other results obtained by using the local spin-density
approximation (LSDA); in particular, the AHC of bcc Fe is in a fair
agreement with the experiment, whereas bigger discrepancies are
encountered for Co and, especially, for Ni. 
This last disagreement has been ascribed to electron correlations,
not treated properly within the LSDA, the effect of which is
particularly strong in Ni \cite{r_2011_fg} and partly also in
Co. \cite{r_2012_tfg}

The calculated Fermi sea term is essentially negligible in fcc Co
and Ni, and it represents a weak effect as compared to the Fermi
surface term in bcc Fe.
However, a completely different picture is obtained for hcp Co,
where the Fermi sea term amounts nearly to 40\% of the total AHC,
irrespective of the orientation of the magnetization, i.e., the
relative anisotropy of the Fermi sea term is similar to that of
the Fermi surface term.
The inclusion of the Fermi sea term brings the present TB-LMTO
results in better agreement with those of the Berry-curvature
approach and with the experiment. \cite{r_2009_rms}
Our calculations prove that the previous statements
\cite{r_2010_nhk, r_2010_lke, r_2012_tkd} about the dominating 
Fermi surface term in metallic systems with high longitudinal
conductivities are not generally valid. 

\subsection{Random alloys and compounds\label{ss_rac}}

\begin{figure}
\includegraphics[width=0.44\textwidth]{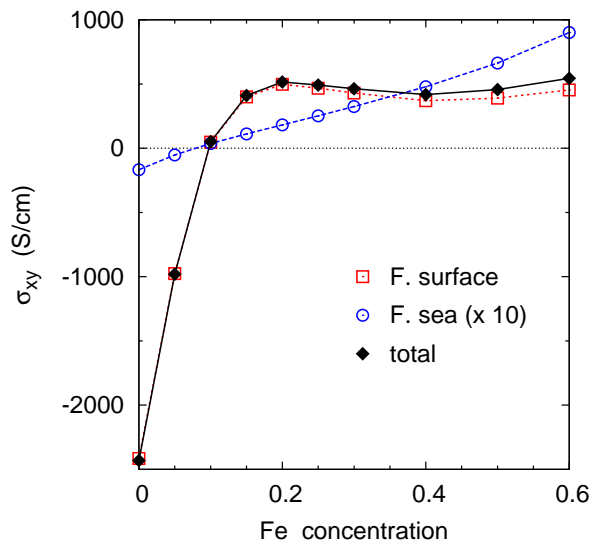}
\caption{(Color online)
The calculated values of the total AHC (full diamonds) and its
Fermi surface (open squares) and Fermi sea (open circles)
contributions in random fcc Ni$_{1-c}$Fe$_c$ alloys as functions
of Fe concentration.
The values of the Fermi sea term are magnified by a factor of 10.
\label{f_nife}}
\end{figure}

An example of the calculated AHC in a concentrated random alloy is
presented in Fig.~\ref{f_nife} for the fcc Ni$_{1-c}$Fe$_c$ system.
One can see that the Fermi sea term represents a very small
correction to the dominating Fermi surface term, as expected from
the similar situation in the pure elements Fe and Ni.
In particular, the previously discussed change of sign of the AHE,
see Ref.~\onlinecite{r_2012_tkd} and references therein, is
encountered at roughly the same concentration.
Note that the Fermi sea term behaves in a smooth manner on the
Ni-rich side despite the strong increase of the total AHC for 
$c \to 0$, in qualitative agreement with the conclusions drawn in
Section \ref{ss_cafst}.

\begin{figure}
\includegraphics[width=0.44\textwidth]{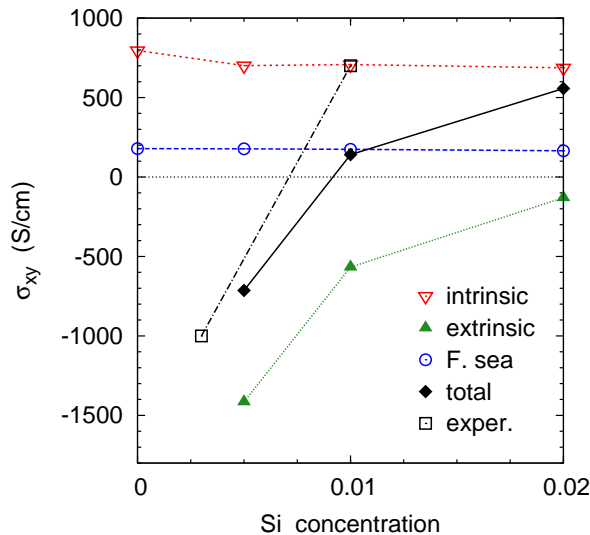}
\caption{(Color online)
The calculated values of the total AHC (full diamonds) and its
intrinsic (open triangles) and extrinsic (full triangles) parts
in diluted bcc Fe$_{1-c}$Si$_c$ alloys as functions of Si
concentration.
The experimental AHC values \cite{r_2009_sot} (open squares) and
the calculated Fermi sea contributions (open circles) are displayed
as well.
\label{f_fesi}}
\end{figure}

Another case study concerns the high-conductivity regime of diluted
bcc Fe$_{1-c}$Si$_c$ alloys, being motivated by recent experiments
\cite{r_2009_sot} for systems with Si impurity concentrations
$c \leq 0.01$. 
The calculated AHCs are displayed in Fig.~\ref{f_fesi} together with
the measured data; the total theoretical values were decomposed
into the intrinsic and extrinsic parts as defined in
Section \ref{ss_tpct}.
One can see a sign change of the AHC in a semiquantitative agreement
with the experiment; this effect can be obviously ascribed to a strong
variation of the extrinsic part, which diverges for $c \to 0$ due to
the skew scattering mechanism, whereas the intrinsic part approaches
smoothly the AHC of pure Fe.
Note that the Fermi sea term, which enters the intrinsic part, is
independent of the Si concentration and it becomes non-negligible 
for compositions with a very small total AHC. 
This system is an example of a ferromagnetic metal containing very
light impurities with a negligible strength of the spin-orbit
interaction and a weak exchange splitting. 
The diverging AHC and the change of its sign in the diluted Fe-Si
alloy prove clearly that such light non-magnetic impurities in a
feromagnetic host with spin-orbit coupling can lead to pronounced
skew-scattering effects in the transverse transport. \cite{r_2010_nso}

\begin{table}
\caption{The calculated values of the AHC for selected ordered and
disordered Co-based Heusler alloys.
The values of the Fermi sea term are displayed in parentheses.
\label{t_cbha}}
\begin{ruledtabular}
\begin{tabular}{lc}
 system  &  $\sigma_{xy}$ (S/cm) \\
\hline
Ideal Co$_2$CrAl & 
                    $400$ ($-107$) \\
(Co$_{0.75}$Cr$_{0.25}$)$_2$(Cr$_{0.5}$Co$_{0.5}$)Al & 
                    $144$ ($39$) \\
Co$_2$(Cr$_{0.75}$Al$_{0.25}$)(Al$_{0.75}$Cr$_{0.25}$) & 
                    $129$ ($4$) \\
Ideal Co$_2$MnAl & 
                    $1787$ ($728$) \\
Co$_2$(Mn$_{0.75}$Al$_{0.25}$)(Al$_{0.75}$Mn$_{0.25}$) & 
                    $452$ ($90$) \\
\end{tabular}
\end{ruledtabular}
\end{table}

The AHE has also been studied intensively in Co-based Heusler alloys
Co$_2$CrAl and Co$_2$MnAl, both experimentally \cite{r_2006_hs,
r_2011_vss} and theoretically. \cite{r_2012_kf, r_2013_tg, r_2013_kdt}
There is a generally accepted view that the structure and chemical
composition of measured samples differ from those of ideal L2$_1$
compounds.
These imperfections are responsible for a discrepancy between the
calculated and measured magnetic moments, strong especially for the
Co$_2$CrAl system, as well as for relatively high longitudinal
resistivities of both systems. 
These facts have partly been explained by an antisite
disorder, \cite{r_2013_kdt} assumed to be of the L2$_1$ type (Co-Cr
swapping) or of the B2 type (Cr-Al swapping) in Co$_2$CrAl, and of
the B2 type (Mn-Al swapping) in Co$_2$MnAl, in agreement with the
experiment in the last case. \cite{r_2011_vss}
Detailed calculations of the electronic structure and of the
AHE, based only on the Fermi surface term, were published in
Ref.~\onlinecite{r_2013_kdt}.
Table \ref{t_cbha} displays the total AHCs for both ideal compounds
and for three disordered systems of compositions
(Co$_{1-c}$Cr$_c$)$_2$(Cr$_{1-2c}$Co$_{2c}$)Al and 
Co$_2$($X_{1-c}$Al$_c$)(Al$_{1-c}X_c$) with $c=0.25$ and $X$ = 
Cr, Mn.
The main conclusions of the previous study, namely, the strong
reduction of the AHC of the ideal compounds by the antisite
disorder and the small extrinsic (vertex corrections) part of
the AHC, \cite{r_2013_kdt} are robust with respect to the inclusion
of the Fermi sea term.
The relative values of the last term lie between 20 and 40 \% of the
total AHC; the only exception is the Co$_2$CrAl alloy with 25\% of
the Cr-Al swap, where the Fermi sea term is essentially negligible.

Let us mention also the inverse Heusler alloy Mn$_2$CoAl, which
represents -- in the ideal structure with a perfect stoichiometry
and without antisite atoms -- an example of the spin-gapless
semiconductor. \cite{r_2013_off}
The AHE of this system at zero temperature is expected to vanish
due to the absence of electron states at the Fermi energy.
This property has been confirmed by recent theoretical calculations
using the Berry-curvature approach \cite{r_2013_off} and the
Kubo-St\v{r}eda formula. \cite{r_2013_kdt}
The present calculation based on the Bastin formula leads to the
Fermi surface and Fermi sea terms as well as to the total AHC
smaller (in absolute values) than $0.5$ S/cm, in very good
agreement with the above theoretical expectation.
Systematic calculations for the system with disorder and further
analysis similar to that in Ref.~\onlinecite{r_2013_kdt} have to
be left for future studies.

\section{Conclusions\label{s_conc}}

We have extended our recent transport theory in the relativistic
TB-LMTO method \cite{r_2012_tkd} by a formulation and numerical
implementation of the Fermi sea term, which follows from the Bastin
formula for the conductivity tensor and which contributes to the AHE.
In the case of random alloys treated in the CPA, the configuration
averaging of this term revealed its purely coherent nature, with
effective vertex corrections originating in the energy dependence
of the average single-particle propagators.
The behavior of the Fermi sea term in the dilute limit of a random
alloy is in general regular, in contrast to the often diverging Fermi
surface term.
We have further examined the transformation properties of the
conductivity tensor with respect to the choice of the LMTO
representation.
This analysis proved the importance of the Fermi sea term for the
representation invariance of the total AHCs and of their intrinsic
part.

The calculations performed with the best-screened LMTO representation
for several qualitatively different systems confirmed in most
cases an expected fact, namely, significantly smaller values of the
Fermi sea term as compared to the Fermi surface term.
Notable exceptions refer to uniaxial systems (hexagonal cobalt)
and to multisublattice multicomponent systems (Heusler alloys). 
However, even in these cases, the inclusion of the Fermi sea term
did not change qualitatively the most important features of the AHE,
such as its anisotropy or sensitivity to antisite defects.
It can be anticipated that the present theory will be useful in
future first-principles studies of transverse transport properties
of prospective materials, in particular with substitutional
disorder.

% If you have acknowledgments, this puts in the proper section head.
\begin{acknowledgments}
The authors acknowledge financial support by the Czech Science
Foundation (Grant No.\ P204/11/1228).
\end{acknowledgments}

% Specify following sections are appendices. Use \appendix* if there
% only one appendix.

\appendix
\section{Energy derivative of the coherent potential
           function\label{app_ed}}

In this Appendix, the formulation of the energy derivative
$\mathcal{P}'(z)$ is briefly sketched, which is based on the
CPA-selfconsistency condition for the coherent potential
function $\mathcal{P}(z)$.
For brevity, the energy argument $z$ of all quantities is
omitted here.

In the single-site CPA, the coherent potential function is
written as a lattice sum over individual sites, $\mathcal{P} 
= \sum_{\bf R} \mathcal{P}_{\bf R}$, and the single-site
contributions $\mathcal{P}_{\bf R}$ are obtained from the vanishing
of the average single-site T-matrix $t_{\bf R}$,
\begin{equation}
\langle t_{\bf R} \rangle = 0 , 
\qquad
t_{\bf R} = 
\left[ 1 + ( P_{\bf R} - \mathcal{P}_{\bf R} ) \bar{g} 
\right]^{-1} 
( P_{\bf R} - \mathcal{P}_{\bf R} ) .
\label{eq_tmat}
\end{equation}
The single-site contributions $\mathcal{P}'_{\bf R}$ to the energy
derivative $\mathcal{P}' = \sum_{\bf R} \mathcal{P}'_{\bf R}$ can be
obtained from the energy derivative of this selfconsistency condition.
This yields:
\begin{eqnarray}
\langle t'_{\bf R} \rangle & = & 0 ,
\nonumber\\
t'_{\bf R} & = & \left[ 1 + 
( P_{\bf R} - \mathcal{P}_{\bf R} ) \bar{g} \right]^{-1}  
( P'_{\bf R} - \mathcal{P}'_{\bf R} ) 
\nonumber\\
 & & + \left[ 1 + ( P_{\bf R} - \mathcal{P}_{\bf R} ) 
\bar{g} \right]^{-1} 
\left[ - ( P_{\bf R} - \mathcal{P}_{\bf R} ) \bar{g}' 
 - ( P'_{\bf R} - \mathcal{P}'_{\bf R} ) \bar{g} \right]
t_{\bf R} 
\nonumber\\
 & = & t_{\bf R} \bar{g} \mathcal{P}' \bar{g} t_{\bf R} 
 + \left[ 1 + ( P_{\bf R} - \mathcal{P}_{\bf R} ) 
\bar{g} \right]^{-1} ( P'_{\bf R} - \mathcal{P}'_{\bf R} )
( 1 - \bar{g} t_{\bf R} ) 
\nonumber\\
 & = & t_{\bf R} \bar{g} \mathcal{P}' \bar{g} t_{\bf R} 
 + ( 1 - t_{\bf R} \bar{g} ) 
( P'_{\bf R} - \mathcal{P}'_{\bf R} )
  ( 1 - \bar{g} t_{\bf R} ) ,
\label{eq_tder}
\end{eqnarray}
where we used the rule (\ref{eq_bgp}) and the identity 
$\left[ 1 + ( P_{\bf R} - \mathcal{P}_{\bf R} ) \bar{g} 
\right]^{-1} = 1 - t_{\bf R} \bar{g}$, which follows from
(\ref{eq_tmat}).
In the first term on the r.h.s., we write explicitly $\mathcal{P}'
= \sum_{{\bf R}'} \mathcal{P}'_{{\bf R}'}$, which leads to
the form
\begin{equation}
 0 = \sum_{{\bf R}'} \langle t_{\bf R} \bar{g} 
\mathcal{P}'_{{\bf R}'} \bar{g} t_{\bf R} \rangle 
  + \langle ( 1 - t_{\bf R} \bar{g} ) 
( P'_{\bf R} - \mathcal{P}'_{\bf R} )
  ( 1 - \bar{g} t_{\bf R} ) \rangle .
\label{eq_work1}
\end{equation}
In the lattice sum, we take out the contribution of the
site ${\bf R}' = {\bf R}$.
After a minor rearrangement of the terms and the use of the
selfconsistency on the ${\bf R}$th site (\ref{eq_tmat}),
we get the final form of the condition for 
$\mathcal{P}'_{\bf R}$:
\begin{equation}
\mathcal{P}'_{\bf R} = 
\langle ( 1 - t_{\bf R} \bar{g} )
 P'_{\bf R} ( 1 - \bar{g} t_{\bf R} ) \rangle 
 + \sum_{{\bf R}' ( \neq {\bf R} )} 
\langle t_{\bf R} \bar{g}
\mathcal{P}'_{{\bf R}'} \bar{g} t_{\bf R} \rangle .
\label{eq_vclike}
\end{equation}
This relation represents a set of coupled linear equations for the
single-site contributions $\mathcal{P}'_{\bf R}$, which is very
similar to that for single-site contributions $\Gamma_{\bf R}$ to
the quantity $\Gamma = \sum_{\bf R} \Gamma_{\bf R}$ relevant for
the general LMTO CPA-vertex corrections. \cite{r_2006_ctk}
Note that the only difference between Eq.~(A5) of 
Ref.~\onlinecite{r_2006_ctk} and the present Eq.~(\ref{eq_vclike})
is in the first term on the r.h.s.
The solution for the $\mathcal{P}'_{\bf R}$ can thus be obtained
by a slightly modified approach described in detail in the Appendix
of Ref.~\onlinecite{r_2006_ctk}.
The energy derivative $P'_{\bf R}$ of the random potential functions
in the first term on the r.h.s.\ of (\ref{eq_vclike}) is calculated
analytically from the parametrized form of the potential functions
$P_{\bf R}$.
The final expression can be written as 
$P'_{\bf R} = {\tilde \mu}_{\bf R} \mu_{\bf R}$, where the
single-site contributions to the random quantities 
$\mu = \sum_{\bf R} \mu_{\bf R}$ and
${\tilde \mu} = \sum_{\bf R} {\tilde \mu}_{\bf R}$ are given by
Eq.~(7) of Ref.~\onlinecite{r_2012_tkd}.

\section{Transformation invariance of the conductivity
         tensor\label{app_tict}}

The study of the invariance of various quantities with respect to the
choice of the LMTO representation is based on relations for the
coherent potential functions and the structure constants in two
different representations, denoted by superscripts $\alpha$ and
$\beta$:
\begin{equation}
\mathcal{P}^\alpha(z) = \left[ 1 + \mathcal{P}^\beta(z)
 (\beta-\alpha) \right]^{-1} \mathcal{P}^\beta(z) , \qquad
S^\alpha = \left[ 1 + S^\beta (\beta-\alpha) \right]^{-1} S^\beta ,
\label{eq_trps}
\end{equation}
where the quantities $\alpha$ and $\beta$ in the brackets denote
non-random site-diagonal matrices of the screening constants.
\cite{r_1997_tdk, r_1986_apj}
Let us abbreviate
\begin{equation}
K = 1 + S^\beta (\beta-\alpha) , \qquad 
K^+ = 1 + (\beta-\alpha) S^\beta , 
\label{eq_defk}
\end{equation}
and $\bar{g}_\pm = \lim_{\epsilon \to 0^+}
\bar{g}(E_\mathrm{F} \pm \mathrm{i} \epsilon)$ 
[and similarly for other energy dependent quantities, such as the
coherent potential functions $\mathcal{P}(z)$ and the
single-site T-matrices $t_{\bf R}(z)$].  
The transformation properties of the effective velocities
(\ref{eq_veloce}) and of the average auxiliary GFs (\ref{eq_avaux})
and their energy derivatives can be summarized as
\begin{eqnarray}
v^\alpha_\mu & = & K^{-1} v^\beta_\mu (K^+)^{-1} , \qquad
\bar{g}^\alpha(z) = K^+ \bar{g}^\beta(z) K + K^+ (\beta-\alpha) ,
\nonumber\\
\bar{g}^\alpha_+ - \bar{g}^\alpha_- & = &  
K^+ ( \bar{g}^\beta_+ - \bar{g}^\beta_- ) K , \qquad
\bar{g}'^{,\alpha}(z) = K^+ \bar{g}'^{,\beta}(z) K , 
\label{eq_trvggg}
\end{eqnarray}
which can be proved by procedures similar to those found in 
Ref.~\onlinecite{r_2012_tkd, r_1986_apj}.

The transformation of the coherent part of the Fermi surface term
(\ref{eq_surf}) is
\begin{equation}
\sigma^{(1),\alpha}_{\mu\nu,\mathrm{coh}} = \sigma_0 \mathrm{Tr}
\left\{ v^\alpha_\mu ( \bar{g}^\alpha_+ - \bar{g}^\alpha_- ) 
v^\alpha_\nu \bar{g}^\alpha_-  -  v^\alpha_\mu \bar{g}^\alpha_+ 
  v^\alpha_\nu ( \bar{g}^\alpha_+ - \bar{g}^\alpha_- ) \right\} 
 = \sigma^{(1),\beta}_{\mu\nu,\mathrm{coh}} + Z_{\mu\nu} ,
\label{eq_tr1coh}
\end{equation}
where the remainder is
\begin{equation}
Z_{\mu\nu} = \sigma_0 \mathrm{Tr}
\left\{ v^\beta_\mu ( \bar{g}^\beta_+ - \bar{g}^\beta_- ) 
v^\beta_\nu (\beta-\alpha) K^{-1} 
 - v^\beta_\mu (\beta-\alpha) K^{-1} 
  v^\beta_\nu ( \bar{g}^\beta_+ - \bar{g}^\beta_- ) \right\} .
\label{eq_zmunu}
\end{equation}
This can be rewritten as
\begin{eqnarray}
Z_{\mu\nu} & = & \sigma_0 \mathrm{Tr}
\left\{ Y_{\mu\nu} ( \bar{g}^\beta_+ - \bar{g}^\beta_- ) \right\} ,
\nonumber\\
Y_{\mu\nu} & = & v^\beta_\mu (\alpha-\beta) K^{-1} v^\beta_\nu
               - v^\beta_\nu (\alpha-\beta) K^{-1} v^\beta_\mu , 
\label{eq_ymunu}
\end{eqnarray}
which proves that for metallic systems, the coherent part of 
$\sigma^{(1)}_{\mu\nu}$ depends on the particular LMTO
representation. 
Note, however, that $Y_{\mu\nu} = - Y_{\nu\mu}$ and 
$Z_{\mu\nu} = - Z_{\nu\mu}$, so that this dependence concerns only
the antisymmetric part of the $\sigma^{(1)}_{\mu\nu,\mathrm{coh}}$
tensor (related to the AHC). 
The symmetric part of $\sigma^{(1)}_{\mu\nu,\mathrm{coh}}$
(related to longitudinal conductivities) is thus invariant with
respect to the choice of the LMTO representation.

The Fermi sea term (\ref{eq_seaq}) transforms as
\begin{equation}
\sigma^{(2),\alpha}_{\mu\nu} = \sigma_0 \int_C \mathrm{d}z  
\mathrm{Tr} \left\{ v^\alpha_\mu \bar{g}'^{,\alpha}(z) v^\alpha_\nu
 \bar{g}^\alpha(z) - v^\alpha_\mu \bar{g}^\alpha(z) v^\alpha_\nu
 \bar{g}'^{,\alpha}(z) \right\} =
\sigma^{(2),\beta}_{\mu\nu} + R_{\mu\nu} ,
\label{eq_tr2}
\end{equation}
where the remainder is
\begin{equation}
R_{\mu\nu} = \sigma_0 \int_C \mathrm{d}z  
\mathrm{Tr} \left\{ v^\beta_\mu \bar{g}'^{,\beta}(z) v^\beta_\nu
 (\beta-\alpha) K^{-1} - v^\beta_\mu (\beta-\alpha) K^{-1}
 v^\beta_\nu \bar{g}'^{,\beta}(z) \right\} .
\label{eq_rmunu}
\end{equation}
This remainder can be rewritten with the use of 
$\int_C \mathrm{d}z \bar{g}'^{,\beta}(z) =
\bar{g}^\beta_- - \bar{g}^\beta_+$.
After a minor rearrangement, we get
\begin{equation}
R_{\mu\nu} = - \sigma_0 \mathrm{Tr}
\left\{ Y_{\mu\nu} ( \bar{g}^\beta_+ - \bar{g}^\beta_- ) \right\} ,
\label{eq_rfin}
\end{equation}
which yields $R_{\mu\nu} + Z_{\mu\nu} = 0$.
This proves that the Fermi sea term alone is sensitive to the
choice of the LMTO representation, but the sum
$\sigma^{(1)}_{\mu\nu,\mathrm{coh}} + \sigma^{(2)}_{\mu\nu}$
is strictly invariant, as mentioned in Section \ref{ss_tpct}.

For the vertex part of the Fermi surface term, transformation
properties are needed for a number of quantities entering the
general expression for the LMTO vertex corrections, \cite{r_2006_ctk}
see also Eq.~(\ref{eq_vg1vgvc}).
For the average auxiliary GFs, it holds
\begin{equation} 
\bar{g}^\alpha(z) = K^+ \bar{g}^\beta(z) 
 \mathcal{P}^\beta(z) \mathcal{P}^{-\alpha}(z) 
= \mathcal{P}^{-\alpha}(z) \mathcal{P}^\beta(z)
  \bar{g}^\beta(z) K ,
\label{eq_trgx}
\end{equation} 
where we abbreviated $\mathcal{P}^{-\alpha}(z) 
= [ \mathcal{P}^\alpha(z) ]^{-1}$.
The quantity $\tilde{g}(z)$ comprising all non-site-diagonal
blocks of $\bar{g}(z)$, i.e.,
$\tilde{g}^{L_1 L_2}_{{\bf R}_1 {\bf R}_2}(z) 
= ( 1 - \delta_{{\bf R}_1 {\bf R}_2} ) 
\bar{g}^{L_1 L_2}_{{\bf R}_1 {\bf R}_2}(z)$,
transforms as
\begin{equation}
\tilde{g}^\alpha(z) = \mathcal{P}^{-\alpha}(z) \mathcal{P}^\beta(z)
\tilde{g}^\beta(z) \mathcal{P}^\beta(z) \mathcal{P}^{-\alpha}(z) .
\label{eq_trnsdg}
\end{equation}
The transformation rule for the quantity 
$\chi^{\Lambda_1 \Lambda_2}_{{\bf R}_1 {\bf R}_2}
= (\tilde{g}_+)^{L_1 L_2}_{{\bf R}_1 {\bf R}_2} 
  (\tilde{g}_-)^{L'_2 L'_1}_{{\bf R}_2 {\bf R}_1}$,
where $\Lambda_1 = (L_1, L'_1)$, $\Lambda_2 = (L_2, L'_2)$,  
follows directly from Eq.~(\ref{eq_trnsdg}).
One obtains
\begin{equation}
\chi^\alpha = \Pi \chi^\beta \tilde{\Pi} ,
\label{eq_trchi}
\end{equation}
where we introduced site-diagonal quantities
$\Pi^{\Lambda_1 \Lambda_2}_{{\bf R}_1 {\bf R}_2} = 
\delta_{{\bf R}_1 {\bf R}_2} \Pi^{\Lambda_1 \Lambda_2}_{{\bf R}_1}$ 
and $\tilde{\Pi}^{\Lambda_1 \Lambda_2}_{{\bf R}_1 {\bf R}_2} 
 = \delta_{{\bf R}_1 {\bf R}_2}
 \tilde{\Pi}^{\Lambda_1 \Lambda_2}_{{\bf R}_1}$, where
\begin{eqnarray}
\Pi^{\Lambda_1 \Lambda_2}_{\bf R} & = &
\left( \mathcal{P}^{-\alpha}_{+,{\bf R}} \mathcal{P}^\beta_{+,{\bf R}} 
\right)^{L_1 L_2}
\left( \mathcal{P}^\beta_{-,{\bf R}} \mathcal{P}^{-\alpha}_{-,{\bf R}} 
\right)^{L'_2 L'_1} , 
\nonumber\\
\tilde{\Pi}^{\Lambda_1 \Lambda_2}_{\bf R} & = &
\left( \mathcal{P}^\beta_{+,{\bf R}} \mathcal{P}^{-\alpha}_{+,{\bf R}} 
\right)^{L_1 L_2}
\left( \mathcal{P}^{-\alpha}_{-,{\bf R}} \mathcal{P}^\beta_{-,{\bf R}} 
\right)^{L'_2 L'_1} .
\label{eq_pitipi} 
\end{eqnarray}
The single-site T-matrices (\ref{eq_tmat}) transform according
to \cite{r_1997_tdk, r_2000_tkd}
\begin{equation}
t^\beta_{\bf R}(z) = \mathcal{P}^\beta_{\bf R}(z)
 \mathcal{P}^{-\alpha}_{\bf R}(z) t^\alpha_{\bf R}(z)
 \mathcal{P}^{-\alpha}_{\bf R}(z) \mathcal{P}^\beta_{\bf R}(z) ,
\label{eq_trtm}
\end{equation}
and the site-diagonal quantity 
$w^{\Lambda_1 \Lambda_2}_{{\bf R}_1 {\bf R}_2} = 
\delta_{{\bf R}_1 {\bf R}_2} 
w^{\Lambda_1 \Lambda_2}_{{\bf R}_1}$, where
$w^{\Lambda_1 \Lambda_2}_{\bf R} = \left\langle 
t^{L_1 L_2}_{+,{\bf R}} t^{L'_2 L'_1}_{-,{\bf R}} \right\rangle$,
satisfies the transformation relation
\begin{equation}
w^\beta = \tilde{\Pi} w^\alpha \Pi .
\label{eq_trw}
\end{equation}
As a consequence of the rules (\ref{eq_trchi}) and (\ref{eq_trw}),
the matrix $\Delta = w^{-1} - \chi$ and its inverse transform as
\begin{equation}
\Delta^\alpha = \Pi \Delta^\beta \tilde{\Pi} , \qquad
(\Delta^\alpha)^{-1} = \tilde{\Pi}^{-1} 
(\Delta^\beta)^{-1} \Pi^{-1} .
\label{eq_trdelin}
\end{equation}
For transformations of the on-site blocks 
$(\bar{g}_+ v_\mu \bar{g}_-)^{L_1 L'_1}_{{\bf R} {\bf R}} \equiv
(\bar{g}_+ v_\mu \bar{g}_-)^{\Lambda_1}_{\bf R}$ and 
$(\bar{g}_- v_\mu \bar{g}_+)^{L'_1 L_1}_{{\bf R} {\bf R}} \equiv 
(\bar{g}_- v_\mu \bar{g}_+)^{\tilde{\Lambda}_1}_{\bf R}$, one
can use the previous relations (\ref{eq_trvggg}) and (\ref{eq_trgx})
for $v_\mu$ and $\bar{g}_\pm$, respectively.
The result is
\begin{equation}
(\bar{g}^\alpha_+ v^\alpha_\mu \bar{g}^\alpha_-)^{\Lambda_1}_{\bf R}
= \sum_{\Lambda_2} \Pi^{\Lambda_1 \Lambda_2}_{\bf R} 
(\bar{g}^\beta_+ v^\beta_\mu \bar{g}^\beta_-)^{\Lambda_2}_{\bf R} ,
\qquad
(\bar{g}^\alpha_- v^\alpha_\mu 
\bar{g}^\alpha_+)^{\tilde{\Lambda}_1}_{\bf R}
= \sum_{\Lambda_2} (\bar{g}^\beta_- v^\beta_\mu 
\bar{g}^\beta_+)^{\tilde{\Lambda}_2}_{\bf R} 
\tilde{\Pi}^{\Lambda_2 \Lambda_1}_{\bf R} , 
\label{eq_trgvg}
\end{equation}
where the $\tilde{\Lambda}_1 = (L'_1, L_1)$
and $\tilde{\Lambda}_2 = (L'_2, L_2)$ denote indices transposed 
to $\Lambda_1 = (L_1, L'_1)$ and $\Lambda_2 = (L_2, L'_2)$, 
respectively. 

The calculation of the vertex part of the Fermi surface term
(\ref{eq_surf}) rests on the formula (\ref{eq_vg1vgvc}).
The identity (\ref{eq_gvgon}) yields
$\mathrm{Tr} \left\langle  v_\mu g_+ 
v_\nu g_+ \right\rangle_\mathrm{VC} =
\mathrm{Tr} \left\langle  v_\mu g_- 
v_\nu g_- \right\rangle_\mathrm{VC} = 0$, 
so that 
$\sigma^{(1)}_{\mu\nu,\mathrm{VC}} = 
2 \sigma_0 \mathrm{Tr} \left\langle  v_\mu g_+ 
v_\nu g_- \right\rangle_\mathrm{VC}$ and
\begin{equation}
\sigma^{(1),\alpha}_{\mu\nu,\mathrm{VC}} = 2 \sigma_0 
\sum_{{\bf R}_1 \Lambda_1} \sum_{{\bf R}_2 \Lambda_2} 
(\bar{g}^\alpha_- v^\alpha_\mu 
\bar{g}^\alpha_+)^{\tilde{\Lambda}_1}_{{\bf R}_1}
\left[ (\Delta^\alpha)^{-1} 
\right]^{\Lambda_1 \Lambda_2}_{{\bf R}_1 {\bf R}_2}
(\bar{g}^\alpha_+ v^\alpha_\nu 
\bar{g}^\alpha_-)^{\Lambda_2}_{{\bf R}_2} .
\label{eq_sig1vc}
\end{equation}
The last relation combined with the transformations
(\ref{eq_trdelin}) and (\ref{eq_trgvg}) leads to the invariance
of the vertex corrections to the Fermi surface term,
$\sigma^{(1),\alpha}_{\mu\nu,\mathrm{VC}}
= \sigma^{(1),\beta}_{\mu\nu,\mathrm{VC}}$.
This completes the proof of the invariance of the total
conductivity tensor $\sigma_{\mu\nu}$ (\ref{eq_sigma}).

% Create the reference section using BibTeX:
%\bibliography{basename of .bib file}
%\bibliography{b_}

\begin{thebibliography}{56}%
\makeatletter
\providecommand \@ifxundefined [1]{%
 \@ifx{#1\undefined}
}%
\providecommand \@ifnum [1]{%
 \ifnum #1\expandafter \@firstoftwo
 \else \expandafter \@secondoftwo
 \fi
}%
\providecommand \@ifx [1]{%
 \ifx #1\expandafter \@firstoftwo
 \else \expandafter \@secondoftwo
 \fi
}%
\providecommand \natexlab [1]{#1}%
\providecommand \enquote  [1]{``#1''}%
\providecommand \bibnamefont  [1]{#1}%
\providecommand \bibfnamefont [1]{#1}%
\providecommand \citenamefont [1]{#1}%
\providecommand \href@noop [0]{\@secondoftwo}%
\providecommand \href [0]{\begingroup \@sanitize@url \@href}%
\providecommand \@href[1]{\@@startlink{#1}\@@href}%
\providecommand \@@href[1]{\endgroup#1\@@endlink}%
\providecommand \@sanitize@url [0]{\catcode `\\12\catcode `\$12\catcode
  `\&12\catcode `\#12\catcode `\^12\catcode `\_12\catcode `\%12\relax}%
\providecommand \@@startlink[1]{}%
\providecommand \@@endlink[0]{}%
\providecommand \url  [0]{\begingroup\@sanitize@url \@url }%
\providecommand \@url [1]{\endgroup\@href {#1}{\urlprefix }}%
\providecommand \urlprefix  [0]{URL }%
\providecommand \Eprint [0]{\href }%
\providecommand \doibase [0]{http://dx.doi.org/}%
\providecommand \selectlanguage [0]{\@gobble}%
\providecommand \bibinfo  [0]{\@secondoftwo}%
\providecommand \bibfield  [0]{\@secondoftwo}%
\providecommand \translation [1]{[#1]}%
\providecommand \BibitemOpen [0]{}%
\providecommand \bibitemStop [0]{}%
\providecommand \bibitemNoStop [0]{.\EOS\space}%
\providecommand \EOS [0]{\spacefactor3000\relax}%
\providecommand \BibitemShut  [1]{\csname bibitem#1\endcsname}%
\let\auto@bib@innerbib\@empty
%</preamble>
\bibitem [{\citenamefont {Strange}(1998)}]{r_1998_ps}%
  \BibitemOpen
  \bibfield  {author} {\bibinfo {author} {\bibfnamefont {P.}~\bibnamefont
  {Strange}},\ }\href@noop {} {\emph {\bibinfo {title} {Relativistic Quantum
  Mechanics}}}\ (\bibinfo  {publisher} {Cambridge University Press},\ \bibinfo
  {year} {1998})\BibitemShut {NoStop}%
\bibitem [{\citenamefont {Hall}(1881)}]{r_1881_eh}%
  \BibitemOpen
  \bibfield  {author} {\bibinfo {author} {\bibfnamefont {E.}~\bibnamefont
  {Hall}},\ }\href@noop {} {\bibfield  {journal} {\bibinfo  {journal} {Philos.
  Mag.}\ }\textbf {\bibinfo {volume} {12}},\ \bibinfo {pages} {157} (\bibinfo
  {year} {1881})}\BibitemShut {NoStop}%
\bibitem [{\citenamefont {Sinitsyn}(2008)}]{r_2008_nas}%
  \BibitemOpen
  \bibfield  {author} {\bibinfo {author} {\bibfnamefont {N.~A.}\ \bibnamefont
  {Sinitsyn}},\ }\href@noop {} {\bibfield  {journal} {\bibinfo  {journal} {J.
  Phys.: Condens. Matter}\ }\textbf {\bibinfo {volume} {20}},\ \bibinfo {pages}
  {023201} (\bibinfo {year} {2008})}\BibitemShut {NoStop}%
\bibitem [{\citenamefont {Nagaosa}\ \emph {et~al.}(2010)\citenamefont
  {Nagaosa}, \citenamefont {Sinova}, \citenamefont {Onoda}, \citenamefont
  {MacDonald},\ and\ \citenamefont {Ong}}]{r_2010_nso}%
  \BibitemOpen
  \bibfield  {author} {\bibinfo {author} {\bibfnamefont {N.}~\bibnamefont
  {Nagaosa}}, \bibinfo {author} {\bibfnamefont {J.}~\bibnamefont {Sinova}},
  \bibinfo {author} {\bibfnamefont {S.}~\bibnamefont {Onoda}}, \bibinfo
  {author} {\bibfnamefont {A.~H.}\ \bibnamefont {MacDonald}}, \ and\ \bibinfo
  {author} {\bibfnamefont {N.~P.}\ \bibnamefont {Ong}},\ }\href@noop {}
  {\bibfield  {journal} {\bibinfo  {journal} {Rev. Mod. Phys.}\ }\textbf
  {\bibinfo {volume} {82}},\ \bibinfo {pages} {1539} (\bibinfo {year}
  {2010})}\BibitemShut {NoStop}%
\bibitem [{\citenamefont {Xiao}\ \emph {et~al.}(2006)\citenamefont {Xiao},
  \citenamefont {Yao}, \citenamefont {Fang},\ and\ \citenamefont
  {Niu}}]{r_2006_xyf}%
  \BibitemOpen
  \bibfield  {author} {\bibinfo {author} {\bibfnamefont {D.}~\bibnamefont
  {Xiao}}, \bibinfo {author} {\bibfnamefont {Y.}~\bibnamefont {Yao}}, \bibinfo
  {author} {\bibfnamefont {Z.}~\bibnamefont {Fang}}, \ and\ \bibinfo {author}
  {\bibfnamefont {Q.}~\bibnamefont {Niu}},\ }\href@noop {} {\bibfield
  {journal} {\bibinfo  {journal} {Phys. Rev. Lett.}\ }\textbf {\bibinfo
  {volume} {97}},\ \bibinfo {pages} {026603} (\bibinfo {year}
  {2006})}\BibitemShut {NoStop}%
\bibitem [{\citenamefont {Onoda}\ \emph {et~al.}(2008)\citenamefont {Onoda},
  \citenamefont {Sugimoto},\ and\ \citenamefont {Nagaosa}}]{r_2008_osn}%
  \BibitemOpen
  \bibfield  {author} {\bibinfo {author} {\bibfnamefont {S.}~\bibnamefont
  {Onoda}}, \bibinfo {author} {\bibfnamefont {N.}~\bibnamefont {Sugimoto}}, \
  and\ \bibinfo {author} {\bibfnamefont {N.}~\bibnamefont {Nagaosa}},\
  }\href@noop {} {\bibfield  {journal} {\bibinfo  {journal} {Phys. Rev. B}\
  }\textbf {\bibinfo {volume} {77}},\ \bibinfo {pages} {165103} (\bibinfo
  {year} {2008})}\BibitemShut {NoStop}%
\bibitem [{\citenamefont {Cr\'epieux}\ and\ \citenamefont
  {Bruno}(2001)}]{r_2001_cb}%
  \BibitemOpen
  \bibfield  {author} {\bibinfo {author} {\bibfnamefont {A.}~\bibnamefont
  {Cr\'epieux}}\ and\ \bibinfo {author} {\bibfnamefont {P.}~\bibnamefont
  {Bruno}},\ }\href@noop {} {\bibfield  {journal} {\bibinfo  {journal} {Phys.
  Rev. B}\ }\textbf {\bibinfo {volume} {64}},\ \bibinfo {pages} {014416}
  (\bibinfo {year} {2001})}\BibitemShut {NoStop}%
\bibitem [{\citenamefont {Karplus}\ and\ \citenamefont
  {Luttinger}(1954)}]{r_1954_kl}%
  \BibitemOpen
  \bibfield  {author} {\bibinfo {author} {\bibfnamefont {R.}~\bibnamefont
  {Karplus}}\ and\ \bibinfo {author} {\bibfnamefont {J.~M.}\ \bibnamefont
  {Luttinger}},\ }\href@noop {} {\bibfield  {journal} {\bibinfo  {journal}
  {Phys. Rev.}\ }\textbf {\bibinfo {volume} {95}},\ \bibinfo {pages} {1154}
  (\bibinfo {year} {1954})}\BibitemShut {NoStop}%
\bibitem [{\citenamefont {Jungwirth}\ \emph {et~al.}(2002)\citenamefont
  {Jungwirth}, \citenamefont {Niu},\ and\ \citenamefont
  {MacDonald}}]{r_2002_jnm}%
  \BibitemOpen
  \bibfield  {author} {\bibinfo {author} {\bibfnamefont {T.}~\bibnamefont
  {Jungwirth}}, \bibinfo {author} {\bibfnamefont {Q.}~\bibnamefont {Niu}}, \
  and\ \bibinfo {author} {\bibfnamefont {A.~H.}\ \bibnamefont {MacDonald}},\
  }\href@noop {} {\bibfield  {journal} {\bibinfo  {journal} {Phys. Rev. Lett.}\
  }\textbf {\bibinfo {volume} {88}},\ \bibinfo {pages} {207208} (\bibinfo
  {year} {2002})}\BibitemShut {NoStop}%
\bibitem [{\citenamefont {Smit}(1955)}]{r_1955_js}%
  \BibitemOpen
  \bibfield  {author} {\bibinfo {author} {\bibfnamefont {J.}~\bibnamefont
  {Smit}},\ }\href@noop {} {\bibfield  {journal} {\bibinfo  {journal}
  {Physica}\ }\textbf {\bibinfo {volume} {21}},\ \bibinfo {pages} {877}
  (\bibinfo {year} {1955})}\BibitemShut {NoStop}%
\bibitem [{\citenamefont {Smit}(1958)}]{r_1958_js}%
  \BibitemOpen
  \bibfield  {author} {\bibinfo {author} {\bibfnamefont {J.}~\bibnamefont
  {Smit}},\ }\href@noop {} {\bibfield  {journal} {\bibinfo  {journal}
  {Physica}\ }\textbf {\bibinfo {volume} {24}},\ \bibinfo {pages} {39}
  (\bibinfo {year} {1958})}\BibitemShut {NoStop}%
\bibitem [{\citenamefont {Berger}(1970)}]{r_1970_lb}%
  \BibitemOpen
  \bibfield  {author} {\bibinfo {author} {\bibfnamefont {L.}~\bibnamefont
  {Berger}},\ }\href@noop {} {\bibfield  {journal} {\bibinfo  {journal} {Phys.
  Rev. B}\ }\textbf {\bibinfo {volume} {2}},\ \bibinfo {pages} {4559} (\bibinfo
  {year} {1970})}\BibitemShut {NoStop}%
\bibitem [{\citenamefont {Yao}\ \emph {et~al.}(2004)\citenamefont {Yao},
  \citenamefont {Kleinman}, \citenamefont {MacDonald}, \citenamefont {Sinova},
  \citenamefont {Jungwirth}, \citenamefont {Wang}, \citenamefont {Wang},\ and\
  \citenamefont {Niu}}]{r_2004_ykm}%
  \BibitemOpen
  \bibfield  {author} {\bibinfo {author} {\bibfnamefont {Y.}~\bibnamefont
  {Yao}}, \bibinfo {author} {\bibfnamefont {L.}~\bibnamefont {Kleinman}},
  \bibinfo {author} {\bibfnamefont {A.~H.}\ \bibnamefont {MacDonald}}, \bibinfo
  {author} {\bibfnamefont {J.}~\bibnamefont {Sinova}}, \bibinfo {author}
  {\bibfnamefont {T.}~\bibnamefont {Jungwirth}}, \bibinfo {author}
  {\bibfnamefont {D.~S.}\ \bibnamefont {Wang}}, \bibinfo {author}
  {\bibfnamefont {E.}~\bibnamefont {Wang}}, \ and\ \bibinfo {author}
  {\bibfnamefont {Q.}~\bibnamefont {Niu}},\ }\href@noop {} {\bibfield
  {journal} {\bibinfo  {journal} {Phys. Rev. Lett.}\ }\textbf {\bibinfo
  {volume} {92}},\ \bibinfo {pages} {037204} (\bibinfo {year}
  {2004})}\BibitemShut {NoStop}%
\bibitem [{\citenamefont {Wang}\ \emph {et~al.}(2007)\citenamefont {Wang},
  \citenamefont {Vanderbilt}, \citenamefont {Yates},\ and\ \citenamefont
  {Souza}}]{r_2007_wvy}%
  \BibitemOpen
  \bibfield  {author} {\bibinfo {author} {\bibfnamefont {X.}~\bibnamefont
  {Wang}}, \bibinfo {author} {\bibfnamefont {D.}~\bibnamefont {Vanderbilt}},
  \bibinfo {author} {\bibfnamefont {J.~R.}\ \bibnamefont {Yates}}, \ and\
  \bibinfo {author} {\bibfnamefont {I.}~\bibnamefont {Souza}},\ }\href@noop {}
  {\bibfield  {journal} {\bibinfo  {journal} {Phys. Rev. B}\ }\textbf {\bibinfo
  {volume} {76}},\ \bibinfo {pages} {195109} (\bibinfo {year}
  {2007})}\BibitemShut {NoStop}%
\bibitem [{\citenamefont {Roman}\ \emph {et~al.}(2009)\citenamefont {Roman},
  \citenamefont {Mokrousov},\ and\ \citenamefont {Souza}}]{r_2009_rms}%
  \BibitemOpen
  \bibfield  {author} {\bibinfo {author} {\bibfnamefont {E.}~\bibnamefont
  {Roman}}, \bibinfo {author} {\bibfnamefont {Y.}~\bibnamefont {Mokrousov}}, \
  and\ \bibinfo {author} {\bibfnamefont {I.}~\bibnamefont {Souza}},\
  }\href@noop {} {\bibfield  {journal} {\bibinfo  {journal} {Phys. Rev. Lett.}\
  }\textbf {\bibinfo {volume} {103}},\ \bibinfo {pages} {097203} (\bibinfo
  {year} {2009})}\BibitemShut {NoStop}%
\bibitem [{\citenamefont {Solovyev}(2003)}]{r_2003_ivs}%
  \BibitemOpen
  \bibfield  {author} {\bibinfo {author} {\bibfnamefont {I.~V.}\ \bibnamefont
  {Solovyev}},\ }\href@noop {} {\bibfield  {journal} {\bibinfo  {journal}
  {Phys. Rev. B}\ }\textbf {\bibinfo {volume} {67}},\ \bibinfo {pages} {174406}
  (\bibinfo {year} {2003})}\BibitemShut {NoStop}%
\bibitem [{\citenamefont {K{\"u}bler}\ and\ \citenamefont
  {Felser}(2012)}]{r_2012_kf}%
  \BibitemOpen
  \bibfield  {author} {\bibinfo {author} {\bibfnamefont {J.}~\bibnamefont
  {K{\"u}bler}}\ and\ \bibinfo {author} {\bibfnamefont {C.}~\bibnamefont
  {Felser}},\ }\href@noop {} {\bibfield  {journal} {\bibinfo  {journal} {Phys.
  Rev. B}\ }\textbf {\bibinfo {volume} {85}},\ \bibinfo {pages} {012405}
  (\bibinfo {year} {2012})}\BibitemShut {NoStop}%
\bibitem [{\citenamefont {Tung}\ and\ \citenamefont {Guo}(2013)}]{r_2013_tg}%
  \BibitemOpen
  \bibfield  {author} {\bibinfo {author} {\bibfnamefont {J.~C.}\ \bibnamefont
  {Tung}}\ and\ \bibinfo {author} {\bibfnamefont {G.~Y.}\ \bibnamefont {Guo}},\
  }\href@noop {} {\bibfield  {journal} {\bibinfo  {journal} {New J. Phys.}\
  }\textbf {\bibinfo {volume} {15}},\ \bibinfo {pages} {033014} (\bibinfo
  {year} {2013})}\BibitemShut {NoStop}%
\bibitem [{\citenamefont {Ouardi}\ \emph {et~al.}(2013)\citenamefont {Ouardi},
  \citenamefont {Fecher}, \citenamefont {Felser},\ and\ \citenamefont
  {K{\"u}bler}}]{r_2013_off}%
  \BibitemOpen
  \bibfield  {author} {\bibinfo {author} {\bibfnamefont {S.}~\bibnamefont
  {Ouardi}}, \bibinfo {author} {\bibfnamefont {G.~H.}\ \bibnamefont {Fecher}},
  \bibinfo {author} {\bibfnamefont {C.}~\bibnamefont {Felser}}, \ and\ \bibinfo
  {author} {\bibfnamefont {J.}~\bibnamefont {K{\"u}bler}},\ }\href@noop {}
  {\bibfield  {journal} {\bibinfo  {journal} {Phys. Rev. Lett.}\ }\textbf
  {\bibinfo {volume} {110}},\ \bibinfo {pages} {100401} (\bibinfo {year}
  {2013})}\BibitemShut {NoStop}%
\bibitem [{\citenamefont {Chen}\ \emph {et~al.}(2014)\citenamefont {Chen},
  \citenamefont {Niu},\ and\ \citenamefont {MacDonald}}]{r_2014_cnm}%
  \BibitemOpen
  \bibfield  {author} {\bibinfo {author} {\bibfnamefont {H.}~\bibnamefont
  {Chen}}, \bibinfo {author} {\bibfnamefont {Q.}~\bibnamefont {Niu}}, \ and\
  \bibinfo {author} {\bibfnamefont {A.~H.}\ \bibnamefont {MacDonald}},\
  }\href@noop {} {\bibfield  {journal} {\bibinfo  {journal} {Phys. Rev. Lett.}\
  }\textbf {\bibinfo {volume} {112}},\ \bibinfo {pages} {017205} (\bibinfo
  {year} {2014})}\BibitemShut {NoStop}%
\bibitem [{\citenamefont {Lowitzer}\ \emph {et~al.}(2010)\citenamefont
  {Lowitzer}, \citenamefont {K{\"o}dderitzsch},\ and\ \citenamefont
  {Ebert}}]{r_2010_lke}%
  \BibitemOpen
  \bibfield  {author} {\bibinfo {author} {\bibfnamefont {S.}~\bibnamefont
  {Lowitzer}}, \bibinfo {author} {\bibfnamefont {D.}~\bibnamefont
  {K{\"o}dderitzsch}}, \ and\ \bibinfo {author} {\bibfnamefont
  {H.}~\bibnamefont {Ebert}},\ }\href@noop {} {\bibfield  {journal} {\bibinfo
  {journal} {Phys. Rev. Lett.}\ }\textbf {\bibinfo {volume} {105}},\ \bibinfo
  {pages} {266604} (\bibinfo {year} {2010})}\BibitemShut {NoStop}%
\bibitem [{\citenamefont {Turek}\ \emph
  {et~al.}(2012{\natexlab{a}})\citenamefont {Turek}, \citenamefont
  {Kudrnovsk\'y},\ and\ \citenamefont {Drchal}}]{r_2012_tkd}%
  \BibitemOpen
  \bibfield  {author} {\bibinfo {author} {\bibfnamefont {I.}~\bibnamefont
  {Turek}}, \bibinfo {author} {\bibfnamefont {J.}~\bibnamefont {Kudrnovsk\'y}},
  \ and\ \bibinfo {author} {\bibfnamefont {V.}~\bibnamefont {Drchal}},\
  }\href@noop {} {\bibfield  {journal} {\bibinfo  {journal} {Phys. Rev. B}\
  }\textbf {\bibinfo {volume} {86}},\ \bibinfo {pages} {014405} (\bibinfo
  {year} {2012}{\natexlab{a}})}\BibitemShut {NoStop}%
\bibitem [{\citenamefont {Kubo}(1957)}]{r_1957_rk}%
  \BibitemOpen
  \bibfield  {author} {\bibinfo {author} {\bibfnamefont {R.}~\bibnamefont
  {Kubo}},\ }\href@noop {} {\bibfield  {journal} {\bibinfo  {journal} {J. Phys.
  Soc. Jpn.}\ }\textbf {\bibinfo {volume} {12}},\ \bibinfo {pages} {570}
  (\bibinfo {year} {1957})}\BibitemShut {NoStop}%
\bibitem [{\citenamefont {Velick\'y}(1969)}]{r_1969_bv}%
  \BibitemOpen
  \bibfield  {author} {\bibinfo {author} {\bibfnamefont {B.}~\bibnamefont
  {Velick\'y}},\ }\href@noop {} {\bibfield  {journal} {\bibinfo  {journal}
  {Phys. Rev.}\ }\textbf {\bibinfo {volume} {184}},\ \bibinfo {pages} {614}
  (\bibinfo {year} {1969})}\BibitemShut {NoStop}%
\bibitem [{\citenamefont {Butler}(1985)}]{r_1985_whb}%
  \BibitemOpen
  \bibfield  {author} {\bibinfo {author} {\bibfnamefont {W.~H.}\ \bibnamefont
  {Butler}},\ }\href@noop {} {\bibfield  {journal} {\bibinfo  {journal} {Phys.
  Rev. B}\ }\textbf {\bibinfo {volume} {31}},\ \bibinfo {pages} {3260}
  (\bibinfo {year} {1985})}\BibitemShut {NoStop}%
\bibitem [{\citenamefont {Carva}\ \emph {et~al.}(2006)\citenamefont {Carva},
  \citenamefont {Turek}, \citenamefont {Kudrnovsk\'y},\ and\ \citenamefont
  {Bengone}}]{r_2006_ctk}%
  \BibitemOpen
  \bibfield  {author} {\bibinfo {author} {\bibfnamefont {K.}~\bibnamefont
  {Carva}}, \bibinfo {author} {\bibfnamefont {I.}~\bibnamefont {Turek}},
  \bibinfo {author} {\bibfnamefont {J.}~\bibnamefont {Kudrnovsk\'y}}, \ and\
  \bibinfo {author} {\bibfnamefont {O.}~\bibnamefont {Bengone}},\ }\href@noop
  {} {\bibfield  {journal} {\bibinfo  {journal} {Phys. Rev. B}\ }\textbf
  {\bibinfo {volume} {73}},\ \bibinfo {pages} {144421} (\bibinfo {year}
  {2006})}\BibitemShut {NoStop}%
\bibitem [{\citenamefont {K{\"o}dderitzsch}\ \emph {et~al.}(2013)\citenamefont
  {K{\"o}dderitzsch}, \citenamefont {Chadova}, \citenamefont {Min\'ar},\ and\
  \citenamefont {Ebert}}]{r_2013_kcm}%
  \BibitemOpen
  \bibfield  {author} {\bibinfo {author} {\bibfnamefont {D.}~\bibnamefont
  {K{\"o}dderitzsch}}, \bibinfo {author} {\bibfnamefont {K.}~\bibnamefont
  {Chadova}}, \bibinfo {author} {\bibfnamefont {J.}~\bibnamefont {Min\'ar}}, \
  and\ \bibinfo {author} {\bibfnamefont {H.}~\bibnamefont {Ebert}},\
  }\href@noop {} {\bibfield  {journal} {\bibinfo  {journal} {New J. Phys.}\
  }\textbf {\bibinfo {volume} {15}},\ \bibinfo {pages} {053009} (\bibinfo
  {year} {2013})}\BibitemShut {NoStop}%
\bibitem [{\citenamefont {Kudrnovsk\'y}\ \emph {et~al.}(2011)\citenamefont
  {Kudrnovsk\'y}, \citenamefont {Drchal}, \citenamefont {Khmelevskyi},\ and\
  \citenamefont {Turek}}]{r_2011_kdk}%
  \BibitemOpen
  \bibfield  {author} {\bibinfo {author} {\bibfnamefont {J.}~\bibnamefont
  {Kudrnovsk\'y}}, \bibinfo {author} {\bibfnamefont {V.}~\bibnamefont
  {Drchal}}, \bibinfo {author} {\bibfnamefont {S.}~\bibnamefont {Khmelevskyi}},
  \ and\ \bibinfo {author} {\bibfnamefont {I.}~\bibnamefont {Turek}},\
  }\href@noop {} {\bibfield  {journal} {\bibinfo  {journal} {Phys. Rev. B}\
  }\textbf {\bibinfo {volume} {84}},\ \bibinfo {pages} {214436} (\bibinfo
  {year} {2011})}\BibitemShut {NoStop}%
\bibitem [{\citenamefont {Kudrnovsk\'y}\ \emph {et~al.}(2013)\citenamefont
  {Kudrnovsk\'y}, \citenamefont {Drchal},\ and\ \citenamefont
  {Turek}}]{r_2013_kdt}%
  \BibitemOpen
  \bibfield  {author} {\bibinfo {author} {\bibfnamefont {J.}~\bibnamefont
  {Kudrnovsk\'y}}, \bibinfo {author} {\bibfnamefont {V.}~\bibnamefont
  {Drchal}}, \ and\ \bibinfo {author} {\bibfnamefont {I.}~\bibnamefont
  {Turek}},\ }\href@noop {} {\bibfield  {journal} {\bibinfo  {journal} {Phys.
  Rev. B}\ }\textbf {\bibinfo {volume} {88}},\ \bibinfo {pages} {014422}
  (\bibinfo {year} {2013})}\BibitemShut {NoStop}%
\bibitem [{\citenamefont {St\v{r}eda}(1982)}]{r_1982_ps}%
  \BibitemOpen
  \bibfield  {author} {\bibinfo {author} {\bibfnamefont {P.}~\bibnamefont
  {St\v{r}eda}},\ }\href@noop {} {\bibfield  {journal} {\bibinfo  {journal} {J.
  Phys. C: Solid State Phys.}\ }\textbf {\bibinfo {volume} {15}},\ \bibinfo
  {pages} {L717} (\bibinfo {year} {1982})}\BibitemShut {NoStop}%
\bibitem [{\citenamefont {Bastin}\ \emph {et~al.}(1971)\citenamefont {Bastin},
  \citenamefont {Lewiner}, \citenamefont {Betbeder-Matibet},\ and\
  \citenamefont {Nozieres}}]{r_1971_blb}%
  \BibitemOpen
  \bibfield  {author} {\bibinfo {author} {\bibfnamefont {A.}~\bibnamefont
  {Bastin}}, \bibinfo {author} {\bibfnamefont {C.}~\bibnamefont {Lewiner}},
  \bibinfo {author} {\bibfnamefont {O.}~\bibnamefont {Betbeder-Matibet}}, \
  and\ \bibinfo {author} {\bibfnamefont {P.}~\bibnamefont {Nozieres}},\
  }\href@noop {} {\bibfield  {journal} {\bibinfo  {journal} {J. Phys. Chem.
  Solids}\ }\textbf {\bibinfo {volume} {32}},\ \bibinfo {pages} {1811}
  (\bibinfo {year} {1971})}\BibitemShut {NoStop}%
\bibitem [{\citenamefont {Kontani}\ \emph {et~al.}(2007)\citenamefont
  {Kontani}, \citenamefont {Tanaka},\ and\ \citenamefont
  {Yamada}}]{r_2007_kty}%
  \BibitemOpen
  \bibfield  {author} {\bibinfo {author} {\bibfnamefont {H.}~\bibnamefont
  {Kontani}}, \bibinfo {author} {\bibfnamefont {T.}~\bibnamefont {Tanaka}}, \
  and\ \bibinfo {author} {\bibfnamefont {K.}~\bibnamefont {Yamada}},\
  }\href@noop {} {\bibfield  {journal} {\bibinfo  {journal} {Phys. Rev. B}\
  }\textbf {\bibinfo {volume} {75}},\ \bibinfo {pages} {184416} (\bibinfo
  {year} {2007})}\BibitemShut {NoStop}%
\bibitem [{\citenamefont {Naito}\ \emph {et~al.}(2010)\citenamefont {Naito},
  \citenamefont {Hirashima},\ and\ \citenamefont {Kontani}}]{r_2010_nhk}%
  \BibitemOpen
  \bibfield  {author} {\bibinfo {author} {\bibfnamefont {T.}~\bibnamefont
  {Naito}}, \bibinfo {author} {\bibfnamefont {D.~S.}\ \bibnamefont
  {Hirashima}}, \ and\ \bibinfo {author} {\bibfnamefont {H.}~\bibnamefont
  {Kontani}},\ }\href@noop {} {\bibfield  {journal} {\bibinfo  {journal} {Phys.
  Rev. B}\ }\textbf {\bibinfo {volume} {81}},\ \bibinfo {pages} {195111}
  (\bibinfo {year} {2010})}\BibitemShut {NoStop}%
\bibitem [{\citenamefont {Turek}\ \emph {et~al.}(2002)\citenamefont {Turek},
  \citenamefont {Kudrnovsk\'y}, \citenamefont {Drchal}, \citenamefont
  {Szunyogh},\ and\ \citenamefont {Weinberger}}]{r_2002_tkd}%
  \BibitemOpen
  \bibfield  {author} {\bibinfo {author} {\bibfnamefont {I.}~\bibnamefont
  {Turek}}, \bibinfo {author} {\bibfnamefont {J.}~\bibnamefont {Kudrnovsk\'y}},
  \bibinfo {author} {\bibfnamefont {V.}~\bibnamefont {Drchal}}, \bibinfo
  {author} {\bibfnamefont {L.}~\bibnamefont {Szunyogh}}, \ and\ \bibinfo
  {author} {\bibfnamefont {P.}~\bibnamefont {Weinberger}},\ }\href@noop {}
  {\bibfield  {journal} {\bibinfo  {journal} {Phys. Rev. B}\ }\textbf {\bibinfo
  {volume} {65}},\ \bibinfo {pages} {125101} (\bibinfo {year}
  {2002})}\BibitemShut {NoStop}%
\bibitem [{\citenamefont {Zabloudil}\ \emph {et~al.}(2005)\citenamefont
  {Zabloudil}, \citenamefont {Hammerling}, \citenamefont {Szunyogh},\ and\
  \citenamefont {Weinberger}}]{r_2005_zhs}%
  \BibitemOpen
  \bibfield  {author} {\bibinfo {author} {\bibfnamefont {J.}~\bibnamefont
  {Zabloudil}}, \bibinfo {author} {\bibfnamefont {R.}~\bibnamefont
  {Hammerling}}, \bibinfo {author} {\bibfnamefont {L.}~\bibnamefont
  {Szunyogh}}, \ and\ \bibinfo {author} {\bibfnamefont {P.}~\bibnamefont
  {Weinberger}},\ }\href@noop {} {\emph {\bibinfo {title} {Electron Scattering
  in Solid Matter}}}\ (\bibinfo  {publisher} {Springer, Berlin},\ \bibinfo
  {year} {2005})\BibitemShut {NoStop}%
\bibitem [{\citenamefont {Ebert}\ \emph {et~al.}(2011)\citenamefont {Ebert},
  \citenamefont {K{\"o}dderitzsch},\ and\ \citenamefont
  {Min\'ar}}]{r_2011_ekm}%
  \BibitemOpen
  \bibfield  {author} {\bibinfo {author} {\bibfnamefont {H.}~\bibnamefont
  {Ebert}}, \bibinfo {author} {\bibfnamefont {D.}~\bibnamefont
  {K{\"o}dderitzsch}}, \ and\ \bibinfo {author} {\bibfnamefont
  {J.}~\bibnamefont {Min\'ar}},\ }\href@noop {} {\bibfield  {journal} {\bibinfo
   {journal} {Rep. Prog. Phys.}\ }\textbf {\bibinfo {volume} {74}},\ \bibinfo
  {pages} {096501} (\bibinfo {year} {2011})}\BibitemShut {NoStop}%
\bibitem [{\citenamefont {Smr\v{c}ka}\ and\ \citenamefont
  {St\v{r}eda}(1977)}]{r_1977_ss}%
  \BibitemOpen
  \bibfield  {author} {\bibinfo {author} {\bibfnamefont {L.}~\bibnamefont
  {Smr\v{c}ka}}\ and\ \bibinfo {author} {\bibfnamefont {P.}~\bibnamefont
  {St\v{r}eda}},\ }\href@noop {} {\bibfield  {journal} {\bibinfo  {journal} {J.
  Phys. C: Solid State Phys.}\ }\textbf {\bibinfo {volume} {10}},\ \bibinfo
  {pages} {2153} (\bibinfo {year} {1977})}\BibitemShut {NoStop}%
\bibitem [{\citenamefont {Solovyev}\ \emph {et~al.}(1991)\citenamefont
  {Solovyev}, \citenamefont {Liechtenstein}, \citenamefont {Gubanov},
  \citenamefont {Antropov},\ and\ \citenamefont {Andersen}}]{r_1991_slg}%
  \BibitemOpen
  \bibfield  {author} {\bibinfo {author} {\bibfnamefont {I.~V.}\ \bibnamefont
  {Solovyev}}, \bibinfo {author} {\bibfnamefont {A.~I.}\ \bibnamefont
  {Liechtenstein}}, \bibinfo {author} {\bibfnamefont {V.~A.}\ \bibnamefont
  {Gubanov}}, \bibinfo {author} {\bibfnamefont {V.~P.}\ \bibnamefont
  {Antropov}}, \ and\ \bibinfo {author} {\bibfnamefont {O.~K.}\ \bibnamefont
  {Andersen}},\ }\href@noop {} {\bibfield  {journal} {\bibinfo  {journal}
  {Phys. Rev. B}\ }\textbf {\bibinfo {volume} {43}},\ \bibinfo {pages} {14414}
  (\bibinfo {year} {1991})}\BibitemShut {NoStop}%
\bibitem [{\citenamefont {Shick}\ \emph {et~al.}(1996)\citenamefont {Shick},
  \citenamefont {Drchal}, \citenamefont {Kudrnovsk\'y},\ and\ \citenamefont
  {Weinberger}}]{r_1996_sdk}%
  \BibitemOpen
  \bibfield  {author} {\bibinfo {author} {\bibfnamefont {A.~B.}\ \bibnamefont
  {Shick}}, \bibinfo {author} {\bibfnamefont {V.}~\bibnamefont {Drchal}},
  \bibinfo {author} {\bibfnamefont {J.}~\bibnamefont {Kudrnovsk\'y}}, \ and\
  \bibinfo {author} {\bibfnamefont {P.}~\bibnamefont {Weinberger}},\
  }\href@noop {} {\bibfield  {journal} {\bibinfo  {journal} {Phys. Rev. B}\
  }\textbf {\bibinfo {volume} {54}},\ \bibinfo {pages} {1610} (\bibinfo {year}
  {1996})}\BibitemShut {NoStop}%
\bibitem [{\citenamefont {Turek}\ \emph {et~al.}(1997)\citenamefont {Turek},
  \citenamefont {Drchal}, \citenamefont {Kudrnovsk\'y}, \citenamefont
  {\v{S}ob},\ and\ \citenamefont {Weinberger}}]{r_1997_tdk}%
  \BibitemOpen
  \bibfield  {author} {\bibinfo {author} {\bibfnamefont {I.}~\bibnamefont
  {Turek}}, \bibinfo {author} {\bibfnamefont {V.}~\bibnamefont {Drchal}},
  \bibinfo {author} {\bibfnamefont {J.}~\bibnamefont {Kudrnovsk\'y}}, \bibinfo
  {author} {\bibfnamefont {M.}~\bibnamefont {\v{S}ob}}, \ and\ \bibinfo
  {author} {\bibfnamefont {P.}~\bibnamefont {Weinberger}},\ }\href@noop {}
  {\emph {\bibinfo {title} {Electronic Structure of Disordered Alloys, Surfaces
  and Interfaces}}}\ (\bibinfo  {publisher} {Kluwer, Boston},\ \bibinfo {year}
  {1997})\BibitemShut {NoStop}%
\bibitem [{\citenamefont {Andersen}\ and\ \citenamefont
  {Jepsen}(1984)}]{r_1984_aj}%
  \BibitemOpen
  \bibfield  {author} {\bibinfo {author} {\bibfnamefont {O.~K.}\ \bibnamefont
  {Andersen}}\ and\ \bibinfo {author} {\bibfnamefont {O.}~\bibnamefont
  {Jepsen}},\ }\href@noop {} {\bibfield  {journal} {\bibinfo  {journal} {Phys.
  Rev. Lett.}\ }\textbf {\bibinfo {volume} {53}},\ \bibinfo {pages} {2571}
  (\bibinfo {year} {1984})}\BibitemShut {NoStop}%
\bibitem [{\citenamefont {Andersen}\ \emph {et~al.}(1986)\citenamefont
  {Andersen}, \citenamefont {Pawlowska},\ and\ \citenamefont
  {Jepsen}}]{r_1986_apj}%
  \BibitemOpen
  \bibfield  {author} {\bibinfo {author} {\bibfnamefont {O.~K.}\ \bibnamefont
  {Andersen}}, \bibinfo {author} {\bibfnamefont {Z.}~\bibnamefont {Pawlowska}},
  \ and\ \bibinfo {author} {\bibfnamefont {O.}~\bibnamefont {Jepsen}},\
  }\href@noop {} {\bibfield  {journal} {\bibinfo  {journal} {Phys. Rev. B}\
  }\textbf {\bibinfo {volume} {34}},\ \bibinfo {pages} {5253} (\bibinfo {year}
  {1986})}\BibitemShut {NoStop}%
\bibitem [{\citenamefont {Kudrnovsk\'y}\ and\ \citenamefont
  {Drchal}(1990)}]{r_1990_kd}%
  \BibitemOpen
  \bibfield  {author} {\bibinfo {author} {\bibfnamefont {J.}~\bibnamefont
  {Kudrnovsk\'y}}\ and\ \bibinfo {author} {\bibfnamefont {V.}~\bibnamefont
  {Drchal}},\ }\href@noop {} {\bibfield  {journal} {\bibinfo  {journal} {Phys.
  Rev. B}\ }\textbf {\bibinfo {volume} {41}},\ \bibinfo {pages} {7515}
  (\bibinfo {year} {1990})}\BibitemShut {NoStop}%
\bibitem [{\citenamefont {Velick\'y}\ \emph {et~al.}(1968)\citenamefont
  {Velick\'y}, \citenamefont {Kirkpatrick},\ and\ \citenamefont
  {Ehrenreich}}]{r_1968_vke}%
  \BibitemOpen
  \bibfield  {author} {\bibinfo {author} {\bibfnamefont {B.}~\bibnamefont
  {Velick\'y}}, \bibinfo {author} {\bibfnamefont {S.}~\bibnamefont
  {Kirkpatrick}}, \ and\ \bibinfo {author} {\bibfnamefont {H.}~\bibnamefont
  {Ehrenreich}},\ }\href@noop {} {\bibfield  {journal} {\bibinfo  {journal}
  {Phys. Rev.}\ }\textbf {\bibinfo {volume} {175}},\ \bibinfo {pages} {747}
  (\bibinfo {year} {1968})}\BibitemShut {NoStop}%
\bibitem [{\citenamefont {Turek}\ \emph {et~al.}(2000)\citenamefont {Turek},
  \citenamefont {Kudrnovsk\'y},\ and\ \citenamefont {Drchal}}]{r_2000_tkd}%
  \BibitemOpen
  \bibfield  {author} {\bibinfo {author} {\bibfnamefont {I.}~\bibnamefont
  {Turek}}, \bibinfo {author} {\bibfnamefont {J.}~\bibnamefont {Kudrnovsk\'y}},
  \ and\ \bibinfo {author} {\bibfnamefont {V.}~\bibnamefont {Drchal}},\ }in\
  \href@noop {} {\emph {\bibinfo {booktitle} {Electronic Structure and Physical
  Properties of Solids}}},\ \bibinfo {series} {Lecture Notes in Physics}, Vol.\
  \bibinfo {volume} {535},\ \bibinfo {editor} {edited by\ \bibinfo {editor}
  {\bibfnamefont {H.}~\bibnamefont {Dreyss\'e}}}\ (\bibinfo  {publisher}
  {Springer, Berlin},\ \bibinfo {year} {2000})\ p.\ \bibinfo {pages}
  {349}\BibitemShut {NoStop}%
\bibitem [{\citenamefont {Turek}\ \emph
  {et~al.}(2012{\natexlab{b}})\citenamefont {Turek}, \citenamefont
  {Kudrnovsk\'y},\ and\ \citenamefont {Carva}}]{r_2012_tkc}%
  \BibitemOpen
  \bibfield  {author} {\bibinfo {author} {\bibfnamefont {I.}~\bibnamefont
  {Turek}}, \bibinfo {author} {\bibfnamefont {J.}~\bibnamefont {Kudrnovsk\'y}},
  \ and\ \bibinfo {author} {\bibfnamefont {K.}~\bibnamefont {Carva}},\
  }\href@noop {} {\bibfield  {journal} {\bibinfo  {journal} {Phys. Rev. B}\
  }\textbf {\bibinfo {volume} {86}},\ \bibinfo {pages} {174430} (\bibinfo
  {year} {2012}{\natexlab{b}})}\BibitemShut {NoStop}%
\bibitem [{\citenamefont {Williams}\ \emph {et~al.}(1982)\citenamefont
  {Williams}, \citenamefont {Feibelman},\ and\ \citenamefont
  {Lang}}]{r_1982_wfl}%
  \BibitemOpen
  \bibfield  {author} {\bibinfo {author} {\bibfnamefont {A.~R.}\ \bibnamefont
  {Williams}}, \bibinfo {author} {\bibfnamefont {P.~J.}\ \bibnamefont
  {Feibelman}}, \ and\ \bibinfo {author} {\bibfnamefont {N.~D.}\ \bibnamefont
  {Lang}},\ }\href@noop {} {\bibfield  {journal} {\bibinfo  {journal} {Phys.
  Rev. B}\ }\textbf {\bibinfo {volume} {26}},\ \bibinfo {pages} {5433}
  (\bibinfo {year} {1982})}\BibitemShut {NoStop}%
\bibitem [{\citenamefont {Zeller}\ \emph {et~al.}(1982)\citenamefont {Zeller},
  \citenamefont {Deutz},\ and\ \citenamefont {Dederichs}}]{r_1982_zdd}%
  \BibitemOpen
  \bibfield  {author} {\bibinfo {author} {\bibfnamefont {R.}~\bibnamefont
  {Zeller}}, \bibinfo {author} {\bibfnamefont {J.}~\bibnamefont {Deutz}}, \
  and\ \bibinfo {author} {\bibfnamefont {P.~H.}\ \bibnamefont {Dederichs}},\
  }\href@noop {} {\bibfield  {journal} {\bibinfo  {journal} {Solid State
  Commun.}\ }\textbf {\bibinfo {volume} {44}},\ \bibinfo {pages} {993}
  (\bibinfo {year} {1982})}\BibitemShut {NoStop}%
\bibitem [{\citenamefont {Dheer}(1967)}]{r_1967_pnd}%
  \BibitemOpen
  \bibfield  {author} {\bibinfo {author} {\bibfnamefont {P.~N.}\ \bibnamefont
  {Dheer}},\ }\href@noop {} {\bibfield  {journal} {\bibinfo  {journal} {Phys.
  Rev.}\ }\textbf {\bibinfo {volume} {156}},\ \bibinfo {pages} {637} (\bibinfo
  {year} {1967})}\BibitemShut {NoStop}%
\bibitem [{\citenamefont {Hou}\ \emph {et~al.}(2012)\citenamefont {Hou},
  \citenamefont {Li}, \citenamefont {Wei}, \citenamefont {Tian}, \citenamefont
  {Wu},\ and\ \citenamefont {Jin}}]{r_2012_hlw}%
  \BibitemOpen
  \bibfield  {author} {\bibinfo {author} {\bibfnamefont {D.}~\bibnamefont
  {Hou}}, \bibinfo {author} {\bibfnamefont {Y.}~\bibnamefont {Li}}, \bibinfo
  {author} {\bibfnamefont {D.}~\bibnamefont {Wei}}, \bibinfo {author}
  {\bibfnamefont {D.}~\bibnamefont {Tian}}, \bibinfo {author} {\bibfnamefont
  {L.}~\bibnamefont {Wu}}, \ and\ \bibinfo {author} {\bibfnamefont
  {X.}~\bibnamefont {Jin}},\ }\href@noop {} {\bibfield  {journal} {\bibinfo
  {journal} {J. Phys.: Condens. Matter}\ }\textbf {\bibinfo {volume} {24}},\
  \bibinfo {pages} {482001} (\bibinfo {year} {2012})}\BibitemShut {NoStop}%
\bibitem [{\citenamefont {Ye}\ \emph {et~al.}(2012)\citenamefont {Ye},
  \citenamefont {Tian}, \citenamefont {Jin},\ and\ \citenamefont
  {Xiao}}]{r_2012_ytj}%
  \BibitemOpen
  \bibfield  {author} {\bibinfo {author} {\bibfnamefont {L.}~\bibnamefont
  {Ye}}, \bibinfo {author} {\bibfnamefont {Y.}~\bibnamefont {Tian}}, \bibinfo
  {author} {\bibfnamefont {X.}~\bibnamefont {Jin}}, \ and\ \bibinfo {author}
  {\bibfnamefont {D.}~\bibnamefont {Xiao}},\ }\href@noop {} {\bibfield
  {journal} {\bibinfo  {journal} {Phys. Rev. B}\ }\textbf {\bibinfo {volume}
  {85}},\ \bibinfo {pages} {220403} (\bibinfo {year} {2012})}\BibitemShut
  {NoStop}%
\bibitem [{\citenamefont {Fuh}\ and\ \citenamefont {Guo}(2011)}]{r_2011_fg}%
  \BibitemOpen
  \bibfield  {author} {\bibinfo {author} {\bibfnamefont {H.~R.}\ \bibnamefont
  {Fuh}}\ and\ \bibinfo {author} {\bibfnamefont {G.~Y.}\ \bibnamefont {Guo}},\
  }\href@noop {} {\bibfield  {journal} {\bibinfo  {journal} {Phys. Rev. B}\
  }\textbf {\bibinfo {volume} {84}},\ \bibinfo {pages} {144427} (\bibinfo
  {year} {2011})}\BibitemShut {NoStop}%
\bibitem [{\citenamefont {Tung}\ \emph {et~al.}(2012)\citenamefont {Tung},
  \citenamefont {Fuh},\ and\ \citenamefont {Guo}}]{r_2012_tfg}%
  \BibitemOpen
  \bibfield  {author} {\bibinfo {author} {\bibfnamefont {J.~C.}\ \bibnamefont
  {Tung}}, \bibinfo {author} {\bibfnamefont {H.~R.}\ \bibnamefont {Fuh}}, \
  and\ \bibinfo {author} {\bibfnamefont {G.~Y.}\ \bibnamefont {Guo}},\
  }\href@noop {} {\bibfield  {journal} {\bibinfo  {journal} {Phys. Rev. B}\
  }\textbf {\bibinfo {volume} {86}},\ \bibinfo {pages} {024435} (\bibinfo
  {year} {2012})}\BibitemShut {NoStop}%
\bibitem [{\citenamefont {Shiomi}\ \emph {et~al.}(2009)\citenamefont {Shiomi},
  \citenamefont {Onose},\ and\ \citenamefont {Tokura}}]{r_2009_sot}%
  \BibitemOpen
  \bibfield  {author} {\bibinfo {author} {\bibfnamefont {Y.}~\bibnamefont
  {Shiomi}}, \bibinfo {author} {\bibfnamefont {Y.}~\bibnamefont {Onose}}, \
  and\ \bibinfo {author} {\bibfnamefont {Y.}~\bibnamefont {Tokura}},\
  }\href@noop {} {\bibfield  {journal} {\bibinfo  {journal} {Phys. Rev. B}\
  }\textbf {\bibinfo {volume} {79}},\ \bibinfo {pages} {100404} (\bibinfo
  {year} {2009})}\BibitemShut {NoStop}%
\bibitem [{\citenamefont {Husmann}\ and\ \citenamefont
  {Singh}(2006)}]{r_2006_hs}%
  \BibitemOpen
  \bibfield  {author} {\bibinfo {author} {\bibfnamefont {A.}~\bibnamefont
  {Husmann}}\ and\ \bibinfo {author} {\bibfnamefont {L.~J.}\ \bibnamefont
  {Singh}},\ }\href@noop {} {\bibfield  {journal} {\bibinfo  {journal} {Phys.
  Rev. B}\ }\textbf {\bibinfo {volume} {73}},\ \bibinfo {pages} {172417}
  (\bibinfo {year} {2006})}\BibitemShut {NoStop}%
\bibitem [{\citenamefont {Vidal}\ \emph {et~al.}(2011)\citenamefont {Vidal},
  \citenamefont {Stryganyuk}, \citenamefont {Schneider}, \citenamefont
  {Felser},\ and\ \citenamefont {Jakob}}]{r_2011_vss}%
  \BibitemOpen
  \bibfield  {author} {\bibinfo {author} {\bibfnamefont {E.~V.}\ \bibnamefont
  {Vidal}}, \bibinfo {author} {\bibfnamefont {G.}~\bibnamefont {Stryganyuk}},
  \bibinfo {author} {\bibfnamefont {H.}~\bibnamefont {Schneider}}, \bibinfo
  {author} {\bibfnamefont {C.}~\bibnamefont {Felser}}, \ and\ \bibinfo {author}
  {\bibfnamefont {G.}~\bibnamefont {Jakob}},\ }\href@noop {} {\bibfield
  {journal} {\bibinfo  {journal} {Appl. Phys. Lett.}\ }\textbf {\bibinfo
  {volume} {99}},\ \bibinfo {pages} {132509} (\bibinfo {year}
  {2011})}\BibitemShut {NoStop}%
\end{thebibliography}

%merlin.mbs apsrev4-1.bst 2010-07-25 4.21a (PWD, AO, DPC) hacked
%Control: key (0)
%Control: author (72) initials jnrlst
%Control: editor formatted (1) identically to author
%Control: production of article title (-1) disabled
%Control: page (0) single
%Control: year (1) truncated
%Control: production of eprint (0) enabled
\providecommand{\noopsort}[1]{}\providecommand{\singleletter}[1]{#1}%

\end{document}